\documentclass[fleqn,usenatbib]{mnras}

\usepackage{newtxtext,newtxmath}

\usepackage[T1]{fontenc}

\DeclareRobustCommand{\VAN}[3]{#2}
\let\VANthebibliography\thebibliography
\def\thebibliography{\DeclareRobustCommand{\VAN}[3]{##3}\VANthebibliography}

\usepackage{graphicx}	
\usepackage{amsmath}	

\title[IXPE Mrk 501: Multiwavelength View]{IXPE Observations of the Blazar Mrk 501 in 2022:\\ 
A Multiwavelength View}

\author[L. Lisalda et al.]{
L. Lisalda,$^{1}$\thanks{E-mail: Lindsey.Lisalda@wustl.edu}
E. Gau,$^{1}$
H. Krawczynski,$^{1}$
F. Tavecchio,$^{2}$
I. Liodakis,$^{3}$
A. Gokus,$^{1}$
N. Rodriguez Cavero,$^{1}$
\newauthor
M. Nowak,$^{1}$
M. Negro,$^{4}$
% ...
\\
Affiliations are shown at the end of the paper
}

\date{Accepted XXX. Received YYY; in original form ZZZ}

\pubyear{2025}

\begin{document}
\label{firstpage}
\pagerange{\pageref{firstpage}--\pageref{lastpage}}
\maketitle

\begin{abstract}
The blazar Markarian 501 (Mrk 501) was observed on three occasions over a 4-month period between 2022 March and 2022 July with the {\it Imaging X-ray Polarimetry Explorer (IXPE)}. 
In this paper, we report for the first time on the third {\it IXPE} observation, performed between 2022 July 9 and 12, during which
{\it IXPE} detected a linear polarization degree of $\Pi_X=6\pm2$ per cent at a polarization angle, measured east of north, of $\Psi_X=143^\circ\pm11^\circ$ within the 2\,--\,8\,keV X-ray band. 
The X-ray polarization angle and degree during this observation are consistent with those obtained during the first two observations.
The chromaticity of the polarization across radio, optical, and X-ray bands is likewise consistent with the result from the simultaneous campaigns during the first two observations. 
Furthermore, we present two types of models to explain the observed spectral energy distributions (SEDs) and energy-resolved polarization: 
a synchrotron self-Compton model with an anisotropic magnetic field probability distribution in the emitting volume, as well as an energy-stratified shock model. 
Our results support both the shock scenario as well as support that small levels of magnetic field anisotropy can explain the observed polarization.
\end{abstract}



\section{Introduction}
\label{s:intro}
Blazars are a class of active galactic nuclei (AGN) characterized by their flux and spectral variability, high polarization, and collimated relativistic jets of plasma. 
The observed emission from such objects arises from the collimated plasma outflow (jet), which emits over a wide range of the electromagnetic spectrum and is oriented at small angles ($\leq10^\circ$) with respect to the line of sight of the observer. 
Multi-wavelength blazar emissions are characterized by broad, double-peaked spectral energy distributions (SEDs). 
Synchrotron emission from relativistic electrons in the jet results in a low-energy component extending from radio to the optical, UV, or X-ray bands, at which point the high-energy component starts to dominate the emission. 
The high-energy peak consists either of photons Compton-scattered from the same electrons producing the synchrotron emission, or of synchrotron emission from secondary particles created by high-energy protons \citep{2022Galax..10..105S}.

Mrk 501 is a nearby ($z=0.034$) high synchrotron peaked (HSP) blazar, with its low-energy emission component peaking in the X-ray band, and its high-energy component peaking in the GeV\,-\,TeV energy range
\citep{1998ApJ...492L..17P,1999A&A...349...11A,2000A&A...353...97K}. The recently-launched {\it Imaging X-ray Polarimetry Explorer} ({\it IXPE}) satellite has allowed 
us to measure, for the first time, the polarization degree and direction of the synchrotron X-rays in the 2\,-\,8 keV 
energy band \citep{ixpe}. These measurements help constrain the uniformity and direction of the magnetic fields experienced by the high-energy electrons emitting the X-rays.

{\it IXPE} observed Mrk 501 three times in 2022. 
The first two observations (March 8\,-\,10, 2022, 100 ksec, and March 26\,-\,28, 2022, 86 ksec) have been reported in a previous paper \citep{liodakis_polarized_2022}.
Those two observations revealed X-ray polarization degrees of $\Pi_X=10\pm2$ per cent and $\Pi_X=11\pm2$ per cent, at electric vector position angles of $\Psi_X=134^\circ\pm5^\circ$ and $\Psi_X=115^\circ\pm4^\circ$, respectively.
\citet{liodakis_polarized_2022} interpreted the strong increase of the linear polarization degree, from $\Pi_O\,\approx\,4$ per cent in the optical band to $\Pi_X\,\approx\,10$ per cent in the X-ray band, as evidence for electron acceleration by shocks. 
In such models, X-rays are emitted by higher-energy electrons closer to the acceleration region, while the optical emission comes from a larger region with a less uniform magnetic field that thereby yields a lower polarization degree for the emission. 
The energy-stratified electron populations can thus explain the observed multiwavelength polarization characteristics of Mrk 501. 
On the other hand, it is possible that models with particle acceleration from magnetic reconnection might lead to faster and/or greater polarization swings in X-rays, especially as compared to energy-stratified shocks, with lower net polarization degrees in the X-ray band owing to averaging over different polarization directions \citep{Zhang_2021,2022A&A...662A..83D}, a result contrary to the observational findings.
However, the results depend on the particular model, and reconnection cannot yet be ruled out.
Finally, as mentioned above, the X-rays from Mrk 501 were found to be polarized in a direction approximately parallel to the jet axis angle of $\Psi_{jet}=120^\circ\pm 12^\circ$ \citep{Weaver2022}.
The polarization direction, too, can be explained by shocks from an energy-stratified model, with the shocks perpendicular to the jet direction and the magnetic field oriented parallel to the shock front.

In this paper, we report on a third {\it IXPE} campaign performed from July 9\,-\,12, 2022, with contemporaneous radio and optical polarization observations and contemporaneous X-ray spectral coverage. 
Similar to the first two observations, this third pointing found the source in a low-flux, or quiescent, state. 
The rest of the paper is organized in the following manner. 
We present the data sets from the three campaigns, including the new data from this campaign, and their analysis methods in Sect.\,\ref{s:data}. Discussion of the light curves, the fitting of X-ray energy and polarization spectra, and the multi-wavelength polarization results can be found in Sect.\,\ref{s:resData}.
We describe our two models of the multi-wavelength spectral and polarization results in Sect.\,\ref{s:model}, and 
report on the results from fitting these two SED models in Sect.\,\ref{s:resModel}, before closing with a discussion of all results in Sect.\,\ref{s:disc}.

\section{Data sets and analysis methods}
\label{s:data}
We combine the {\it IXPE} observations with campaigns covering various bands of the electromagnetic spectrum. During the first two {\it IXPE} observations of Mrk 501, several other observatories provided contemporaneous coverage in radio, infrared, optical, ultraviolet, and X-rays. 
For the observations, data, and analysis of those two March observations, which are similar to those added here from the third epoch, we refer the reader to \citep{liodakis_polarized_2022}. 
We describe below the bands of coverage for the third (July) observation, along with the processing and analysis methods used.

\subsection{{\it IXPE} X-ray Observations}
The new {\it IXPE} Mrk 501 observations were taken between 2022-07-09 23:35 UTC and 2022-07-12 00:40 UTC.
In all, 97 ksec of spectropolarimetric data were acquired. 
A circular region of radius 1.5 arcminutes and an annulus region with radii 2.5 and 4.5 arcminutes, centered on the source, were used for extracting the source and background counts, respectively.

In order to obtain for this paper the overall polarization degree and angle measurements for each of the three observations, the {\it ixpeobssim} software suite \citep{2022SoftX..1901194B} was used to combine and visualise the data from the source region files of the three {\it IXPE} detectors and yield model-free polarization results.
The background counts were subtracted from the source counts before the spectropolarimetric fits were done in {\it Sherpa} \citep{2001SPIE.4477...76F}. 
\\[1ex]
\subsection{Additional X-ray Observations}
We also use data from the XRT onboard the {\it Neil Gehrels Swift} observatory \citep{gehrels2004}, reduced using the {\it xrtpipeline} (version 0.13.7). 
The {\it Swift}-XRT observations presented in this paper cover a time range from the beginning of 2022 until the end of August 2022, and were performed in ``window timing mode''.

We furthermore include two {\it XMM-Newton} \citep{jansen2001} observations taken on 22 and 24 March 2022. 
For the data extraction, we have used the standard methods with XMMSAS (version 20.0.0). 
In our analysis, we only consider the EPIC-pn data, which was taken in ``Small Window Mode''. 
We extracted the source from a circular region centered on the source coordinates (RA = $253.4675867\degr$, Dec.\ = $39.7600296\degr$), with a radius of $35''$. 
As pile-up was detected, we excluded an inner circular region with a radius of $5''$ from our source region. 
For the background region, we extracted a circular region centered on (RA = $253.4846759\degr$, Dec.\ = 39.8063927$\degr$), with a radius of $60''$, as far from the source as possible. 

The hard X-ray range presented in this work is covered by four observations with the Nuclear Spectroscopic Telescope Array \citep[\textit{NuSTAR};][]{harrison2013}. 
We extracted the data of both Focal Plane Modules A and B (FPMA and FPMB) with nupipeline (version 0.4.9), {\textsc HEAsoft} (version 6.31.1), and {\textsc CALDB} (version 20230504). 
The source region was a circular region centered on (RA = $253.4657281\degr$, Dec.\ = $39.7601030\degr$), with a radius of $80''$. 
The background region was a circular region centered on (RA = $253.5406350\degr$, Dec.\ = $39.8665981\degr$), with a radius of $160''$.

The X-ray energy spectral fits of the {\it Swift}-XRT observations (year-long) and the joint {\it XMM}, {\it Swift}, and {\it NuSTAR} observations (March 2022) were performed with an absorbed log-parabolic model. 
The neutral hydrogen column density was set to the Galactic value of $N_{\mathrm{H}}\,=\,1.69\times 10^{20}\,\mathrm{cm}^{-2}$ \citep{2016A&A...594A.116H}.
We used \texttt{Vern} cross-sections \citep{1996ApJ...465..487V} and the \texttt{wilm} abundances \citep{2000ApJ...542..914W}.
The {\it Swift}-XRT observations were 
fit from 0.5 to 10 keV.
The {\it NuSTAR} FPMA and FPMB data were fit with the spectral parameters $\alpha$, $\beta$, and flux tied together, while accounting for the normalisation difference via a constant left free in the fit.
The best-fit results of the simultaneous observations and of the {\it Swift} data alone are given in Tables~\ref{tab:mrk501_nustar_logparfits} and \ref{tab:mrk501_swift_logparfits1} (with Tables \ref{tab:mrk501_swift_logparfits2} and \ref{tab:mrk501_swift_logparfits3} being continuations of the latter) in the Appendix, respectively.
\\[1ex]
\subsection{Radio Observations}
For the radio observations, we used the Owens Valley Radio Observatory (15~GHz) 40m telescope and the Mets\"{a}hovi Radio Observatory (22 and 37~GHz) 13.7m diameter telescope, as well as radio polarization measurements from the IRAM-30m, taken at 86~GHz and 230~GHz on 11 July (MJD~59771.755).
The OVRO observation and data reduction procedures are described in detail in \cite{Richards2011}. For the Mets\"{a}hovi observations, the telescope detection limit at 37 GHz is $\sim0.2$ Jy under optimal conditions. We consider all data points with a signal-to-noise ratio $<$\,4$\sigma$ as non-detections. 
The errors in the flux density are estimated by accounting for both the contribution of the measurement root-mean-square (RMS) and the uncertainty in the absolute flux calibration.
A detailed description of the data reduction and analysis can be found in \cite{Teraesranta1998}. 

We obtained millimeter-wave polarimetric observations at the Pico Veleta Observatory (Sierra Nevada, Granada, Spain) using the IRAM 30 m telescope. 
The observations were taken at 3.5 mm (86.24 GHz) and 1.3 mm (230 GHz) as part of the POLAMI (Polarimetric Monitoring of AGN at Millimeter Wavelengths\footnote{http://polami.iaa.es/}) program \citep{Agudo2018a, Agudo2018b, Thum2018}. 
In this observing setup, we simultaneously recorded the four (\textit{I, Q, U, V}) Stokes parameters, using the XPOL procedure \citep{Thum2008}. 
A more detailed description of the data reduction, calibration, and managing and flagging procedures can be found in \cite{Agudo2018a}.

Additional mm-radio observations at 225.5~GHz were obtained using the SubMillimeter Array (SMA) \cite{Ho2004} as part of the SMA Monitoring of AGN with POLarization (SMAPOL) program on the 2022 July 10 (MJD~59770.46). The SMA observations are taken using the SMA polarimeter \cite{Marrone2008} and SWARM correlator \cite{Primiani2016}  in full polarization mode. The polarization information for the source is extracted from the Stokes \textit{I, Q, U} visibilities and the data are then calibrated using the MIR\footnote{\url{https://lweb.cfa.harvard.edu/~cqi/mircook.html} software package \citep{Saul1995}}.
\\[1ex]
\subsection{Infrared, Optical, and UV Observations} 
In the optical band, the Burst Optical Observer and Transient Exploring System (BOOTES) supplied $g,r,i$ photometry.
A description of the observations can be found in \cite{Castro-Tirado2012} and \cite{Hu2023}. The Perkins Telescope, Calar Alto, St. Petersburg, NOT, T60, BOOTES, and KANATA observatories provided data in units of magnitudes, which were converted to fluxes for the SEDs via the appropriate Johnsons-Cousins filter (or, for BOOTES, the SDSS filter) conversions. 
We also included optical and UV observations from {\it Swift}-UVOT. 

Optical (R-band) polarization measurements
were taken with the Nordic Optical Telescope (NOT) and RoboPol polarimeter at the Skinakas observatory.
The polarization results are corrected for the unpolarized host-galaxy emission \citep{Nilsson2007,Hovatta2016}. 
Further details on the data acquisition and reduction for these observatories can be found in \citet{liodakis_polarized_2022,DiGesu2022-Mrk421,Middei2023}. 
\\[1ex]
\begin{figure*}
\centering
\includegraphics[width=16cm]{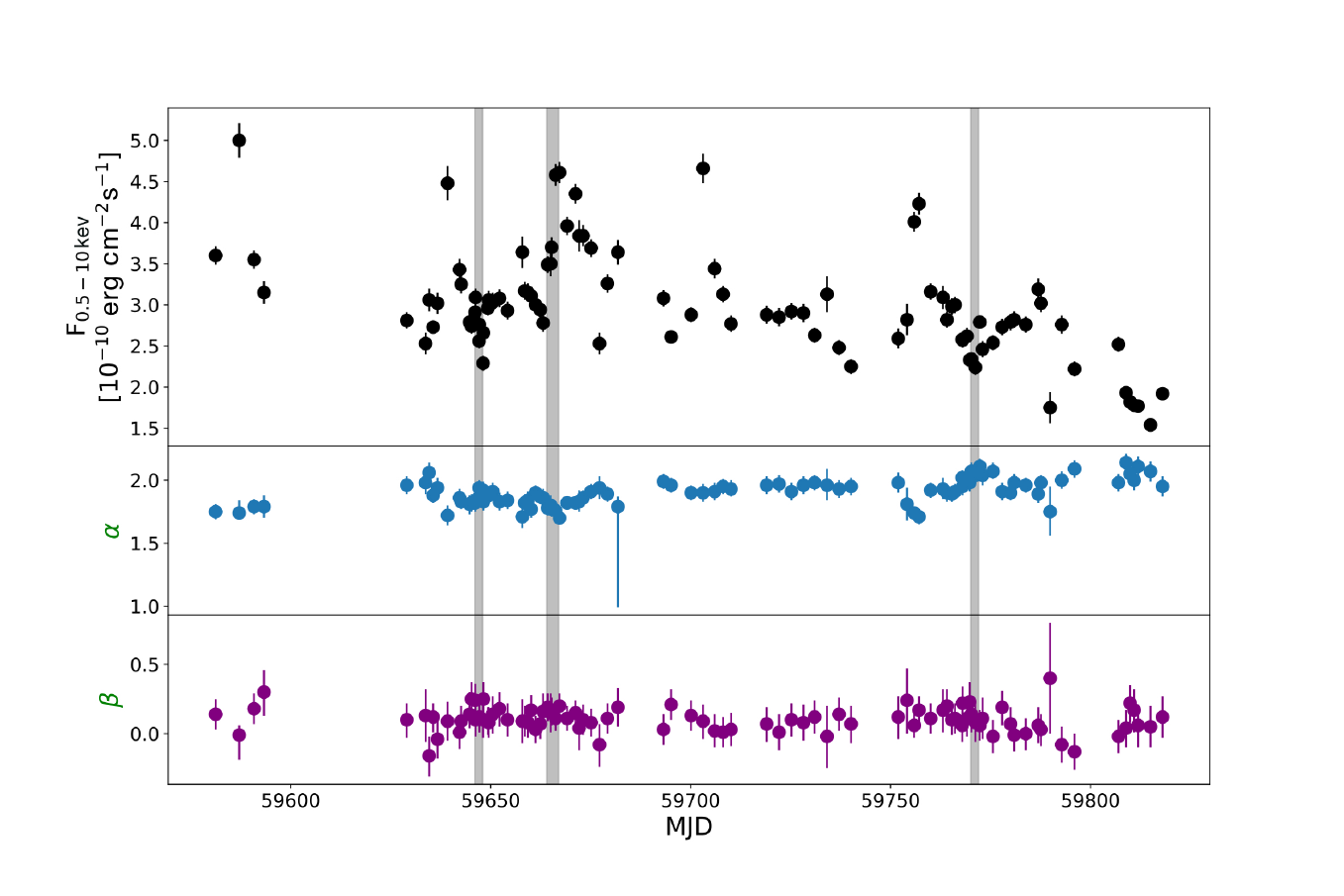}
\caption{Fluxes (upper panel) and spectral parameters $\alpha$ (middle panel) and $\beta$ (lower panel) retrieved via modelling the {\it Swift}-XRT Mrk 501 observations from January to August 2022 with a log-parabola model. 
The grey vertical bands show the times of the three {\it IXPE} observations campaigns. Detailed results from fitting the {\it XMM-Newton}, {\it NuSTAR}, and {\it Swift}-XRT data are found in Tables \ref{tab:mrk501_swift_logparfits1} - \ref{tab:mrk501_swift_logparfits3} in the Appendix.}
\label{fig:xrt}
\end{figure*}
\subsection{Gamma-ray Observations}
The {\it IXPE} observations were performed during a period in which the $\gamma$-ray flux of  Mrk 501 was near its average.  
Hence, we use the time-averaged {\it Fermi}-LAT 4FGL spectrum as a proxy for a simultaneous GeV energy spectrum \citep{2020ApJS..247...33A}.

\section{Observational Results}
\label{s:resData}
\subsection{X-ray Light Curves}
\begin{figure*}
\includegraphics[width=\textwidth]{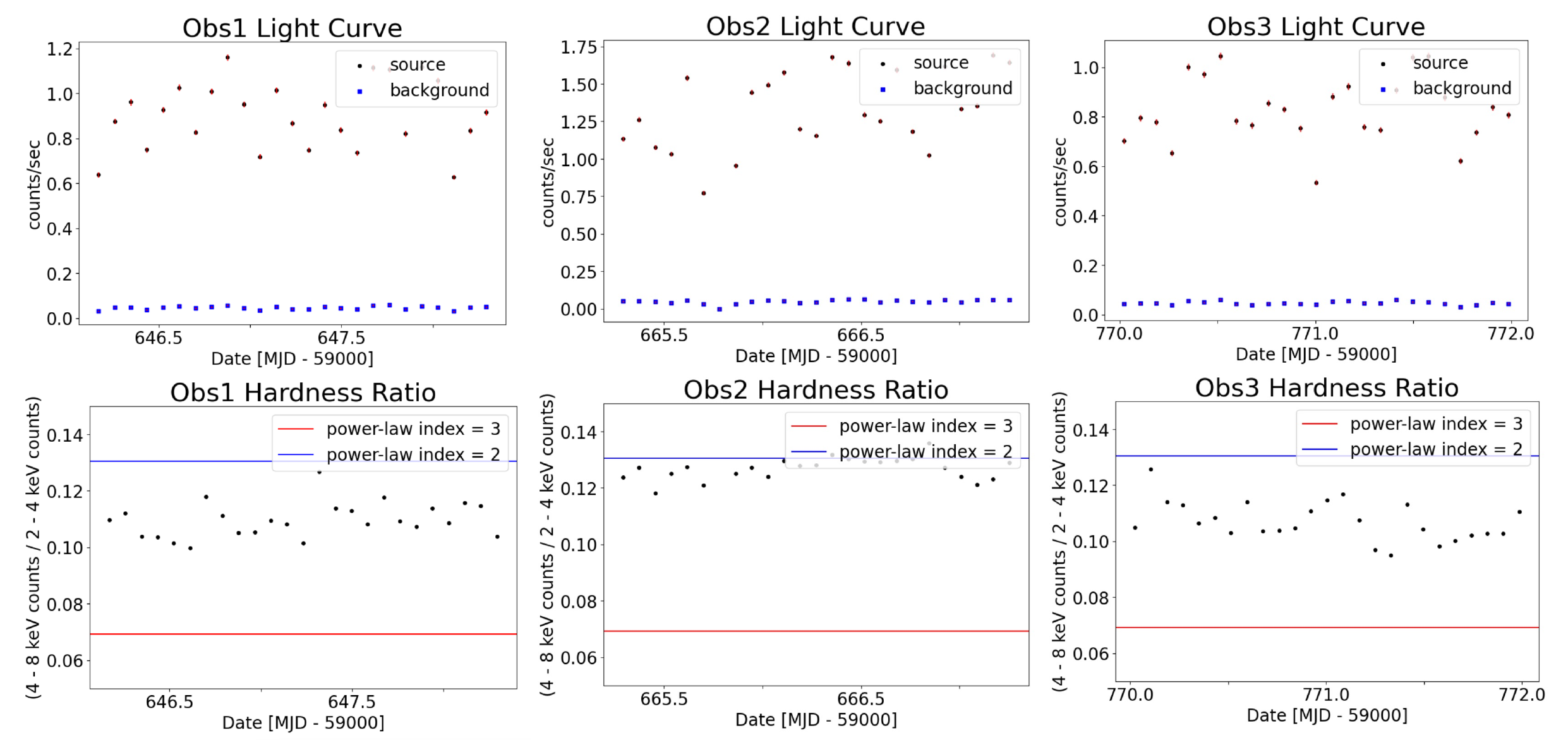}\caption{Light curves and 4--8\,keV/2--4\,keV hardness ratios for each of the three {\it IXPE} campaigns.
The first observation has 7647 seconds per bin for both the light curve and hardness ratios; the second, 7060 seconds per bin; the third, 7069 seconds per bin. 
A more coarse-grained time-binning of the counts in the second observation will reveal a similar rise in flux to that shown in the {\it SWIFT}-XRT light curve of Figure \ref{fig:xrt}.
The (higher up) blue line gives the ratio of 4--8\,keV/2--4\,keV counts that would be obtained by {\it IXPE}, as calculated by {\it WEBPIMMS}, if the spectrum consisted solely of a power-law component with index 2. 
Likewise, the red line gives the ratio, if the spectrum consisted solely of a power-law component with index 3 (i.e., a softer spectrum).}
 \label{fig:LcHardness}
\end{figure*}

Figure \ref{fig:xrt} displays the {\it Swift}-XRT X-ray light curve ($0.5-10$\,keV) from January to August 2022. We obtained the flux values, shown in the upper panel, by fitting a log-parabola model to each spectrum. We display the best-fit spectral parameters, the photon index $\alpha$ at 1 keV and the spectral curvature $\beta$, in the two lower panels.
The flux of Mrk\,501 varied by a factor of up to three during the monitoring coverage of eight months. During some of the times when the source shows a rising flux, the spectrum becomes significantly harder (e.g., at MJD $\sim$59640, $\sim$5965, $\sim$59755), but not every time (e.g., MJD $\sim$59705).
The curvature of the X-ray spectrum slightly varies over time, too, but with less of an apparent connection to the flux state.
As can be seen from the X-ray monitoring with {\it Swift}-XRT, the first and third {\it IXPE} observations caught the source at a similar, rather low flux state. 
The second {\it IXPE} observation occurred during a fast flux increase. 
Figure\,\ref{fig:LcHardness} shows the light curves, as well as the 4--8\,keV/2--4\,keV hardness ratios, for each of the three {\it IXPE} three observations of Mrk 501. 
The higher average flux during the second observation accompanies a corresponding average increase in the hardness ratios (and thus harder energy spectra), the same correlation as exhibited in earlier campaigns \citep[e.g.,][]{2000A&A...353...97K}.

\subsection{Spectropolarimetric Results}

\begin{figure*}
\includegraphics[width=\textwidth]{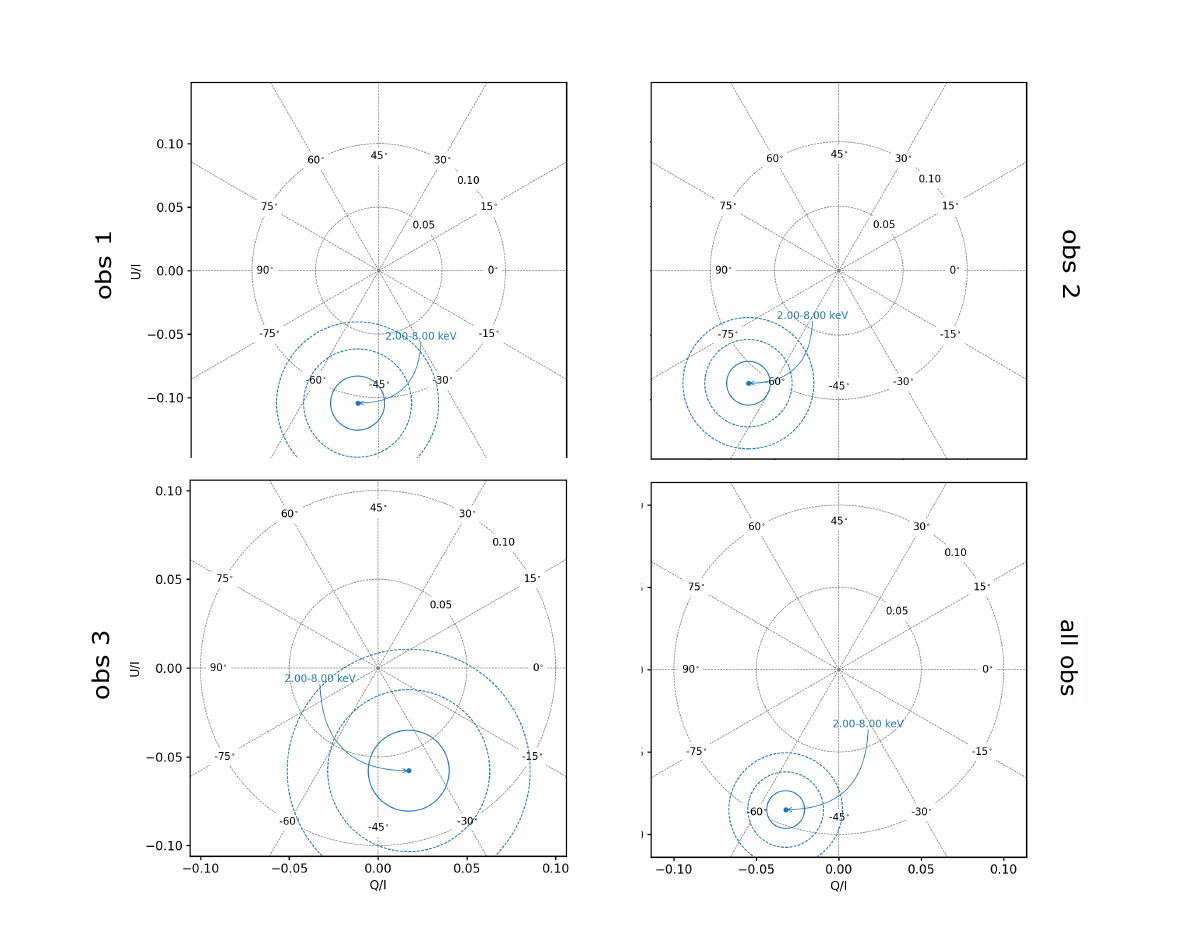}
\caption{Detected Stokes Q and U for the three {\it IXPE} observations of Mrk 501, as well as for all three observations taken together.
The circles around the result display, from innermost to outermost, the 1, 2, and 3$\sigma$ uncertainty regions. 
Note that a polarization angle of $135^\circ$ is equivalent to $-45^\circ$.}
 \label{fig:PCUBE}
\end{figure*}

\begin{table*}
\centering
\caption{Summary of the X-ray polarization results from the model-free analysis enabled by the {\it PCUBE} algorithm of the {\it ixpeobssim} software \citep{2022SoftX..1901194B}.
Results are given for the three {\it IXPE} observations, as well as for combining the data from all three observations, just as is shown in Figure \ref{fig:PCUBE}. 
Errors are given at the $1\sigma$ confidence level. 
The lower significance of the third observation is mainly a result of the polarization degree being measured to be of a lower magnitude than in the first two observations.
Note that a polarization angle of $135^\circ$ is also equivalent to $-45^\circ$.}
\label{tab:pcube_results}
    \begin{tabular}{lcccc} 
		\hline
		Parameter & Obs 1 & Obs 2 & Obs 3 & All Obs \\
		\hline
		Stokes {\it I} & 5133\,$\pm$\,24 & 7712$\pm$\,30 & 4332\,$\pm$\,22 & 17177\,$\pm$\,44 \\
            Stokes {\it Q/I} & $-$0.017\,$\pm$\,0.021 & $-$0.070\,$\pm$\,0.017 & 0.017\,$\pm$\,0.023 & $-$0.032\,$\pm$\,0.011 \\
            Stokes {\it U/I} & $-$0.104\,$\pm$\,0.021 & $-$0.087\,$\pm$\,0.017 & $-$0.058\,$\pm$\,0.023 & $-$0.085\,$\pm$\,0.011 \\
            PD & 0.106\,$\pm$\,0.021 & 0.112\,$\pm$\,0.017 & 0.060\,$\pm$\,0.023 & 0.091\,$\pm$\,0.011 \\
            PA & $-$50\,$\pm$\,6 & $-$64\,$\pm$\,4 & $-$37\,$\pm$\,11 & $-$55\,$\pm$\,4 \\
            Significance $\left[\sigma\right]$ & 4.43 & 6.17 & 1.88 & 7.53 \\
		\hline
	\end{tabular}
\end{table*}

Figure \ref{fig:PCUBE} and Table\,\ref{tab:pcube_results} present the polarimetric results from each of the three {\it IXPE} observations, as well as the combined results from all three observations, produced with the {\it PCUBE} algorithm from {\it ixpeobssim} \citep{2022SoftX..1901194B}.
Polarization was detected at the $>4\,\sigma$--level in observations 1 and 2 and detected at close to the 2\,$\sigma$--level in the third observation. 
For the third observation, a linear polarization degree of $6.0\pm2.3$ per cent and a polarization angle of $143^{\circ}\pm11^{\circ}$ were found. 
Overall, we do not find significant evidence for the X-ray polarization changing between the three observations. 
Although the third observation, when taken by itself, seems to indicate a polarization angle more discrepant from the jet angle, the result obtained from the combination of all three observations remains quite consistent with the previous finding of alignment between the jet ($\rm \psi_{jet}=120^\circ\pm12^\circ$)  
and X-ray polarization ($\psi_{X}=125^\circ\pm4^\circ$) angles\footnote{https://www.bu.edu/blazars/VLBA\_GLAST/1652.html}, with the combined polarization detection being at an even higher significance of 7.5$\sigma$.

\begin{figure*}
\centering
\includegraphics[width=14cm]{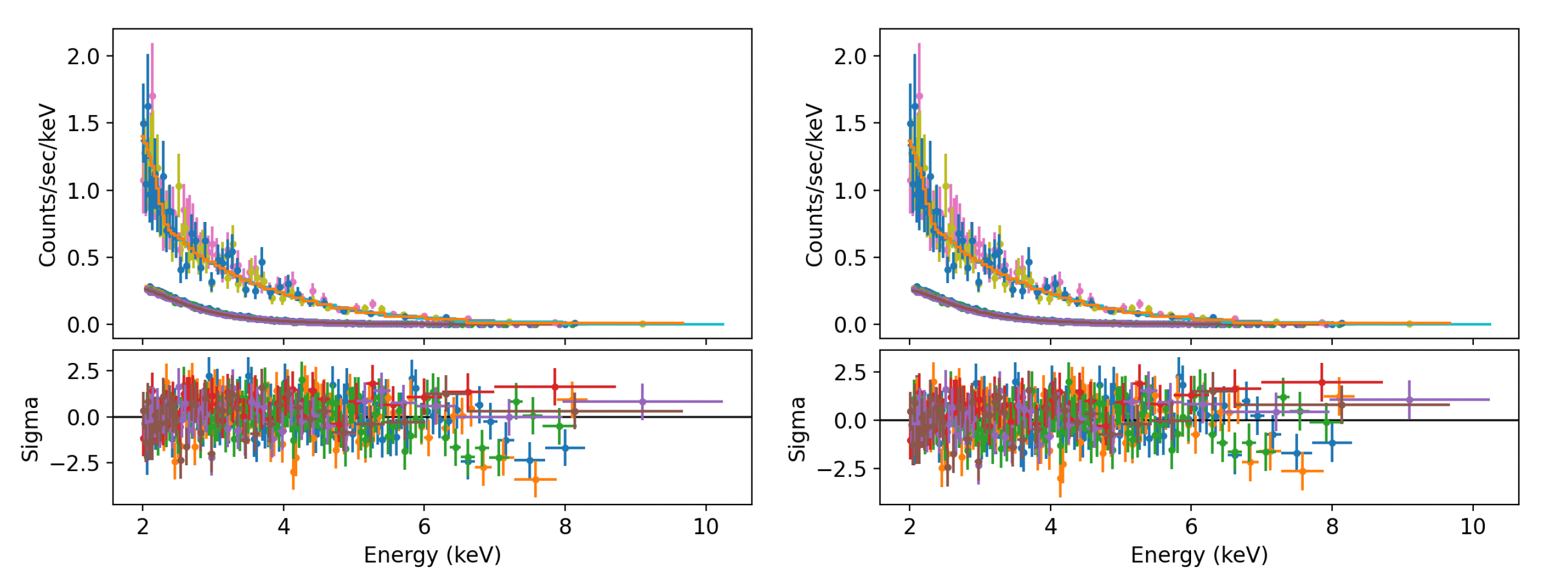}\caption{{\it SWIFT}-XRT and {\it IXPE} data for Stokes I, fitted with an absorbed power-law model (left) and an absorbed log-parabola model (right). 
The data have been re-binned to have a minimum signal-to-noise ratio of 5 in each bin. 
The lower flux blue-green-purple data points, with error bars, are those of the three {\it IXPE} detectors from the third {\it IXPE} observation of Mrk 501. 
The higher flux blue-pink-yellow data points, with error bars, are those of the three {\it SWIFT} observations concurrent with the third {\it IXPE} observation. The solid lines are the spectral models, obtained with the \textit{Sherpa} fitting software, that best explain each data set. The residuals are given in the subplots below the main figures, with the y-axis being in units of the error of a given data point.}\label{fig:SpectroPolarimetricFits}
\end{figure*}

\begin{table*}
\centering
\caption{Results from the joint spectropolarimetric fits of all of the {\it IXPE} and {\it Swift}-XRT data during the time of the third {\it IXPE} observation. 
The hydrogen absorption $\rm n_H$ is fixed to $\rm 0.0155\times10^{22}\,$cm$^{-2}$. 
The pivot energy of the log-parabola model was set to 2 \,keV. 
The instrumental cross-calibration factors are normalized such that the {\it IXPE} DU1 factor is always unity.
Errors are given at the $1\sigma$ confidence level.}
	\label{tab:SpectropolarimetricFits}
	\begin{tabular}{lccc} 
		\hline
		Model Parameter & Absorbed Power-Law & Absorbed Log-Parabola \\
		\hline
             {\it IXPE} DU1 & 1 (frozen) & 1 (frozen) \\
             {\it IXPE} DU2 & 0.950\,$\pm$\,0.009 & 0.950\,$\pm$\,0.009 \\
             {\it IXPE} DU3 & 0.892\,$\pm$\,0.008 & 0.892\,$\pm$\,0.008 \\
             SWIFT & 1.15\,$_{-0.02}^{+0.03}$ & 1.15\,$\pm$\,0.03 \\
             power-law index & 2.41\,$_{-0.02}^{+0.01}$ & n/a \\
             power-law norm [at 1 keV] & 0.067\,$\pm$\,0.001 & n/a \\
             log-parabola alpha & n/a & 2.20\,$_{-0.05}^{+0.04}$ \\
             log-parabola beta & n/a & 0.5\,$\pm$\,0.1 \\
             log-parabola norm & n/a & 0.0124\,$\pm$\,0.0001 \\
             polarization degree (\%) & 7\,$\pm$\,2 & 7\,$\pm$\,2 \\
             polarization angle (deg) & 135\,$\pm$\,8 & 135\,$\pm$\,8 \\
             $\chi^2$ for spectral parameters & 399 & 375 \\
             d.o.f. for spectral parameters & 425 & 424 \\
             $\chi^2$ for polarization parameters & 186 & 185 \\
             d.o.f. for polarization parameters & 178 & 178 \\
		\hline
	\end{tabular}
\end{table*}
We then carried out joint spectropolarimetric fits for the third {\it IXPE} observation and the three {\it SWIFT}-XRT observations that occurred on the same UTC calendar days. 
The data for all three Stokes parameters were background-subtracted before fitting. 
The data were fit to two separate models: an absorbed power-law spectrum, as well as an absorbed log-parabola model, both commonly used for describing blazar X-ray SEDs. 
A constant polarization model (one polarization degree and one polarization angle applicable throughout the 2\,-\,8 keV range) was applied to each of the models. 
The results are given in Table \ref{tab:SpectropolarimetricFits}, and the spectral fits are plotted in Figure \ref{fig:SpectroPolarimetricFits}.
The fitting of the two models produces very similar (and reasonable) $\chi^2$ goodness-of-fit statistics, even though the log-parabolic model has one more fitting parameter.
Moreover, both models give polarization results for the third {\it IXPE} observation consistent with each other and with the spectral, model-free analysis: a polarization degree of 7$\pm$2 per cent and a polarization angle of 135$^{\circ}\pm8^{\circ}$.

For the duration of the third {\it IXPE} observation, we have obtained contemporaneous polarization measurements in the radio and optical bands, as described in Sect.~\ref{s:data}.
The radio emission exhibits a polarization degree of $\rm \Pi_R=1.51\pm0.33$ per cent along a radio polarization angle of $\rm \psi_R=144^\circ\pm6^\circ$ at 86~GHz.
In the mm-radio range, the IRAM-30m observations at 230 GHz yielded only an upper limit of $\Pi_R < 4.6$ per cent (with a 99 per cent confidence). At 225.5~GHz, using SMA, the source shows $\Pi_R=1.27\pm0.28$ per cent along $\psi_R=130\pm5^\circ$.
Thus, the radio measurements of the polarization angle are consistent with those from the X-rays.
The host-galaxy-corrected \citep{Nilsson2007,Hovatta2016} optical (R-band) polarization degree from NOT and RoboPol during the same time period corresponds to $\rm \Pi_O=2.43\pm0.53$ per cent along $\rm \psi_O=113^\circ\pm8^\circ$. 
\\[1ex]

\begin{figure}
\vspace*{0.5cm}
\includegraphics[width=\columnwidth]{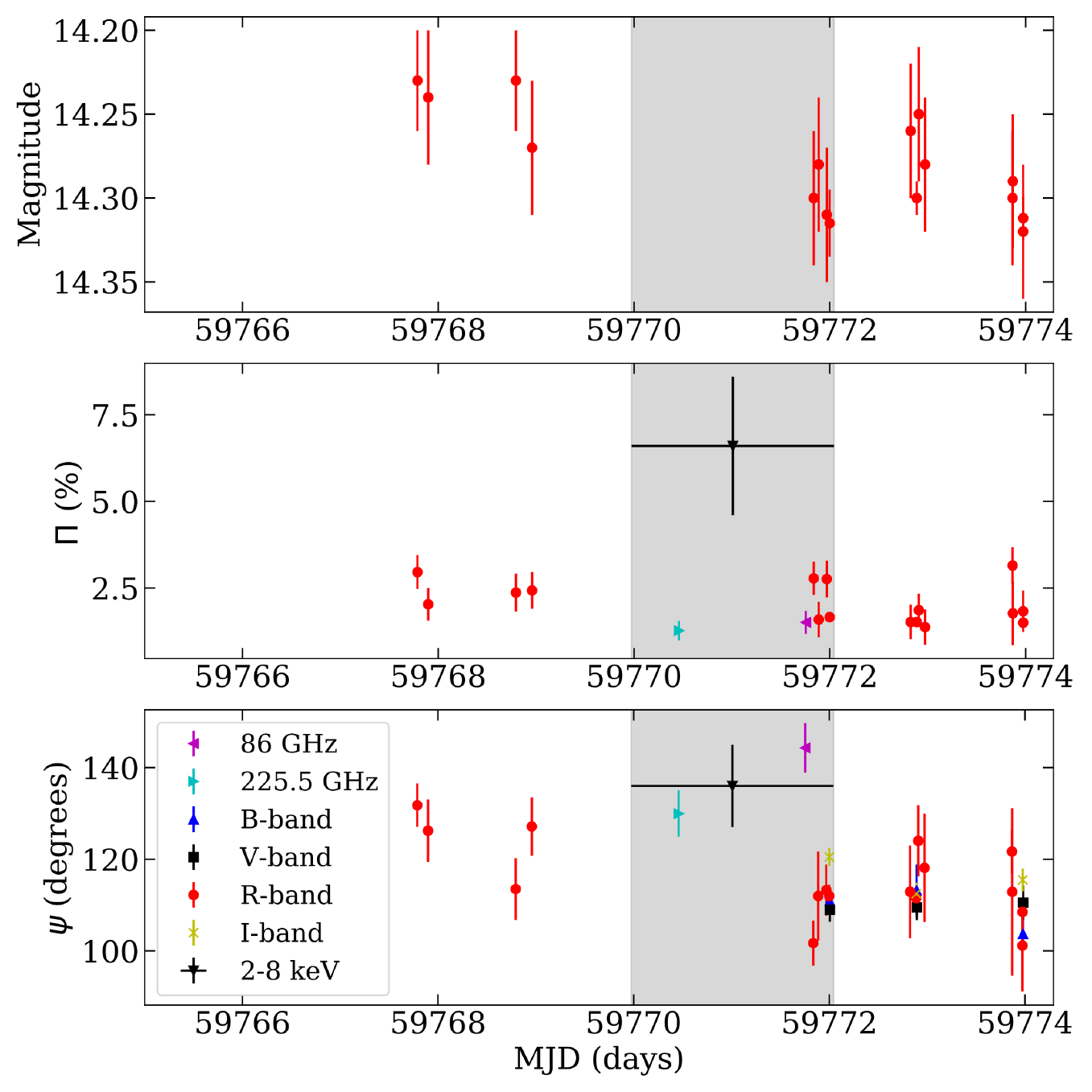}\caption{Contemporaneous radio, optical, and X-ray polarization observations around the time of the third {\it IXPE} observation. 
The top panel shows the R-band magnitudes, the middle panel the measured polarization degrees, and the bottom panel the measured polarization angles. 
The grey-shaded area marks the duration of the third {\it IXPE} observation.}\label{fig:third_obs_pol}
\end{figure}
Figure \ref{fig:third_obs_pol} summarizes the multiwavelength polarization results obtained during the time of the third {\it IXPE} observation. 
The X-ray polarization degree is about a factor of three higher than that detected in the optical band and a factor of four higher than that detected in the radio band. 
The multi-wavelength polarization results from this campaign are consistent with those from the previous two, featuring a strongly chromatic polarization degree that is roughly aligned with the jet axis on the plane of the sky at $\rm \psi_{jet}=120^\circ\pm12^\circ$.
\section{Leptonic Emission Models for the Spectral Energy Distributions}\label{s:model}
In leptonic emission models, the low-energy component is produced by synchrotron emission from electrons and possibly positrons, while the high-energy component is produced by electrons inverse-Compton-scattering synchrotron photons or other photons (e.g., accretion disk photons).
We focus here on synchrotron self-Compton \citep{Jones1974} (SSC) models which have successfully been used to fit blazar energy spectra (see \cite{boettcher2010models} for a review).
The first model is a one-zone model with a non-uniform magnetic field. The increase of the polarization from the optical to the X-ray bands implies that the electrons emitting the optical (X-ray) emission see a less (more) uniform magnetic field distribution. 
The electrons emitting in the optical have lower energies, lose their energy on longer time scales, and can therefore travel through regions with greater variation of magnetic field direction. 
In the shock-in-jet model, we model the evolution of the magnetic field in the downstream medium, and use the magnetic field evolution to predict the polarization degrees.

\subsection{Synchrotron Self-Compton Model with Anisotropic Magnetic Field}
One of the two models that we use to describe the data uses the SSC code described in \citep{2004ApJ...601..151K}.
The SSC code assumes a spherical emission volume of radius $R$ traveling with velocity $\beta c$ ($c$ being the speed of light) along the jet axis. 
The observer views the emission volume at an observer-frame angle $\theta_{\rm obs}$ away from the jet axis.
According to the standard equations of special relativity, the relativistic Doppler factor is given by the equation
\begin{equation}
\delta\,=\, \left[\Gamma(1-\beta\cos{\theta_{\rm obs}})\right]^{-1}.
\label{e:delta}
\end{equation} 
The cosine of the angle to the jet axis
$\mu_{\rm obs}\equiv\cos{\theta_{\rm obs}}$ transforms into the plasma frame according to
\begin{equation}
\mu_{\rm j}\,=\,
\frac{\mu_{\rm obs}-\beta}{1-\mu_{\rm obs}\beta}.
\end{equation} 
An isotropic distribution of electrons emits synchrotron radiation in magnetic fields having strength $B$. 
The synchrotron emission scattering off the electrons gives rise to the high-energy emission component. 
The electron energy spectrum is described by a broken power law with index $p_1$ ($dN_{\rm e}/d\gamma\propto \gamma^{-p_1}$) from electron 
Lorentz factor $\gamma_{\rm min}$ to $\gamma_{\rm b}$ and with index $p_2$ from $\gamma_{\rm b}$
to $\gamma_{\rm max}$.

\begin{figure}
\includegraphics[width=\columnwidth]{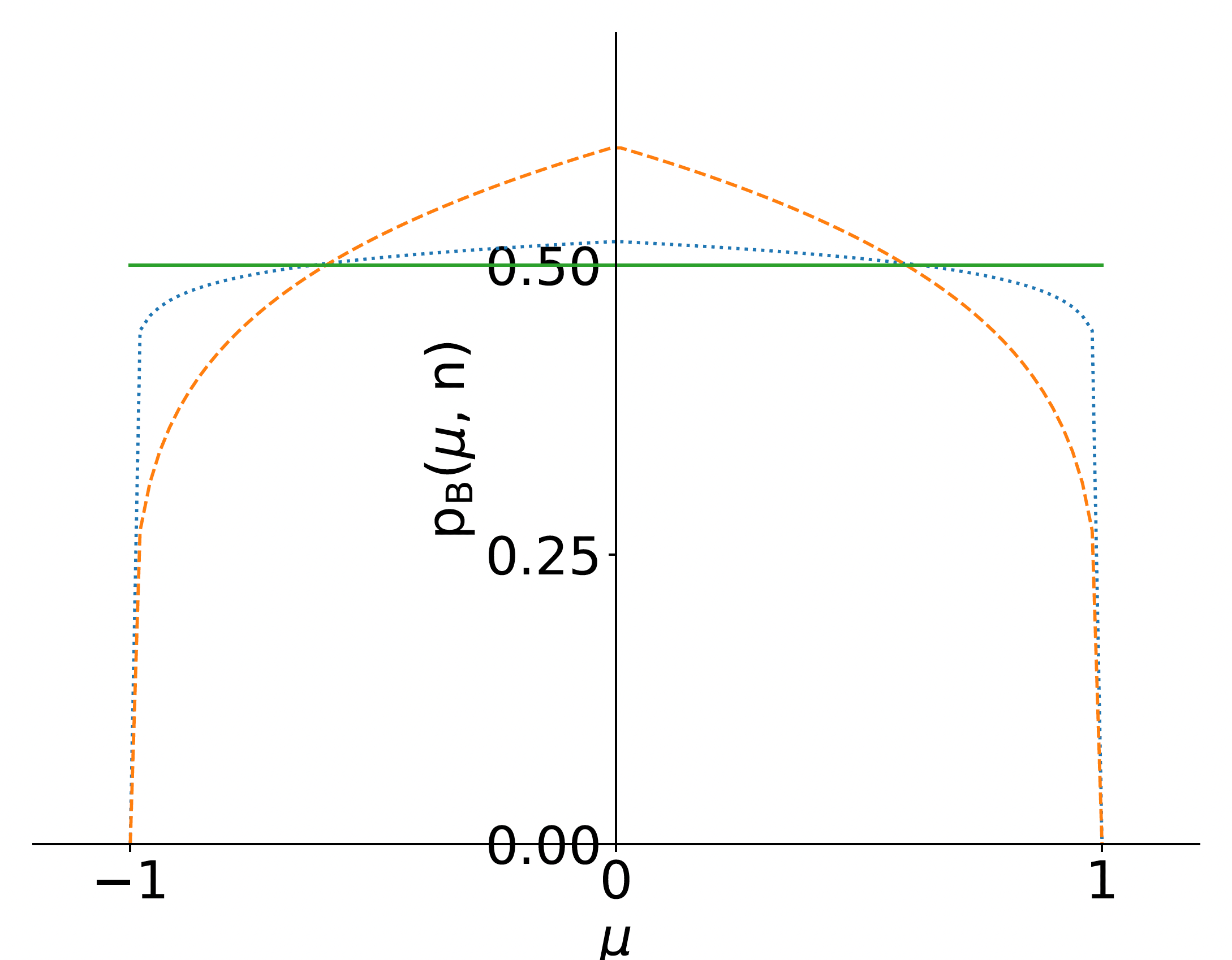}
\caption{We use a simple phenomenological prescription to study the impact of anisotropies in the magnetic field distribution. We assume that the overall distribution of magnetic field directions in all minuscule sub-volumes of the sphere can be described by a probability distribution $p_{\rm B}(\mu,n)$ depending on the cosine $\mu$ of the angle between the magnetic field direction within that sub-volume and the jet axis, with the parameter $n$ characterizing the amount of anisotropy in the overall distribution. We show here the probability distributions for $n=0.041$ (dotted blue line) and $n=0.205$ (dashed orange line), which can explain the optical and X-ray polarization degree results, respectively, reported in Section \ref{s:resData}. The two distributions demonstrate that small deviations from a magnetic field distribution having all directions equally likely are sufficient to explain the varied observed polarization degrees. The flat (isotropic) distribution of $p_{\rm B}=1/2$ is shown by the green solid line.
\label{fig:probs}}
\end{figure}

For this work, we have added the ability to predict the polarization of the synchrotron emission. 
To do so, we assume that the magnetic field in the emission region has the same strength everywhere in the spherical emission volume, but that the magnetic field direction varies locally between each minuscule sub-volume of the sphere. 
The myriad of small sub-volumes thus has each exhibiting its own magnetic field direction, not necessarily related to the direction of the field in an adjacent sub-volume. 
The condition imposed on these directions, however, is that the overall distribution of magnetic field directions (when looking across all sub-volumes) exhibits an anisotropy that can be described with an axisymmetric probability distribution $p_{\rm B}(\mu)$ depending only on the cosine $\mu$ of the angle between the magnetic field direction within the sub-volume and the jet axis (with all angles in the jet frame) for a given \textit{n}.
\textit{n}, then, is used to provide a simple parameterization:
\begin{equation}
p_{\rm B}(\mu,n) =
\frac{n+1}{2}
(1-|\mu|)^n.
\label{e:prob}
\end{equation}
This prescription thus yields an isotropic distribution of magnetic fields for $n=0$ and a distribution of magnetic fields all perpendicular to the jet axis for large $n$, and is normalized to give $\int_{-1}^{1}\,d\mu\, p_{\rm B}(\mu,n)\,=\,1$. 
Figure\,\ref{fig:probs} shows the two sample probability distributions used later to fit the observed polarization degrees.

We calculate the emitted synchrotron Stokes $I$ and Stokes $Q$ parameters as a function of the emitted frequency by integrating over the contributions from all electron Lorentz factors and all magnetic field orientations \citep[see][for a similar calculation]{laing02}. 
In the jet frame, the emitted synchrotron power per angular frequency $\omega$ is given by
\begin{equation} \label{eq4}
\begin{split}
I\,=\,
&\int_{\gamma_{\rm min}}^ {\gamma_{\rm max}} \,d\gamma\,
\frac{dN_{\rm e}}{d\gamma}
\int_{-1}^ {1} d\mu\, \\
&\int_{0}^ {2\pi} d\phi\,
p_{\rm B}(\mu,n)
\frac{\sqrt{3}Bq^3\sin{\alpha}}{4\pi^{2}mc^{2}}
F(x).
\end{split}
\end{equation}
The first integral runs over the Lorentz factors $\gamma$ of the electrons, and the second and third integrals run over the magnetic field directions (i.e., over $\mu$, which is the cosine of the angle between the magnetic field direction and the jet axis, as well as over the azimuthal angle $\phi$). 
$q$ is the charge of the electron, $B$ is the magnetic field strength, $\alpha$ is the angle between the magnetic field direction and the line of sight, and $m$ is the electron mass. 
The variable $x$ depends on all of the integration variables and is given by $x=\omega/\omega_{\rm c}$, with the synchrotron critical frequency $\omega_{\rm c}=3\gamma^2qB\sin{\alpha}/2mc$.
The function $F(x)$ \citep{1986rpa..book.....R} is an integral of the modified Bessel function of order 5/3:
\[
F(x)\,=\,x \int_x^{\infty} K_{5/3}(\xi) d\xi.
\]

The corresponding power in Stokes $Q$ parallel to the jet axis projected on the plane of the sky is given by
\begin{equation} \label{eq5}
\begin{split}
Q\,=\,
\int_{\gamma_{\rm min}}^ {\gamma_{\rm max}} &\,d\gamma\,
\frac{dN_{\rm e}}{d\gamma}
\int_{-1}^ {1} d\mu\,
\int_{0}^ {2\pi} d\phi\,
p_{\rm B}(\mu,n) \,\times\\
&\frac{\sqrt{3}Bq^3\sin{\alpha}}{4\pi^{2} mc^{2}}
G(x)
\cos{2\chi}.
\end{split}
\end{equation}
The terms on the right hand side of Equation (\ref{eq5}) up to the $\cos{2\,\chi}$ term give the intensity of the linearly polarized synchrotron emission.
The factor $\alpha$, in particular, accounts for the dependency of the intensity on the angle between the magnetic field direction and the direction of the line of sight. 
The function $G(x)$ \citep{1986rpa..book.....R} is a function of the modified Bessel function of order 2/3:
\[
G(x)\,=\,x \,K_{2/3}(x).
\]
$\chi$ is the angle between the electric field polarization direction (perpendicular to the magnetic field) and the jet axis projected onto the plane of the sky.
Thus, the factor $\cos{2\,\chi}$ extracts the $Q$-component of the intensity \citep{1986rpa..book.....R} where,
in our treatment, $Q>0$ denotes the linear polarization parallel to the jet axis, and $Q<0$ denotes linear polarization perpendicular to the jet axis. 
The azimuthal symmetry of the distribution of magnetic fields, $p_{\rm B}$, implies that $p_{\rm B}$ is ultimately anisotropic only with respect to the angle of declination from the jet axis (with no anisotropies with respect to the azimuthal angle around the jet axis). 
Thus, after summing the contributions from all azimuthal angles, the net polarization can only be parallel to or perpendicular to the jet axis, and
Stokes $U$ (denoting the plasma-frame intensity of the emission linearly polarized along the diagonals between the jet axis and the directions perpendicular to it) must thus vanish.
The polarization degree can then be given just by $|Q|/I$: the polarization angle is, by symmetry, either 0$^{\circ}$ ($Q>0$) or 90$^{\circ}$ ($Q<0$) with respect to the jet axis.

Our code integrates over all directions of the 4$\pi$ sphere. 
For each direction, we calculate $\mu$ and $\chi$ with the help of simple vector calculus equations. Each result contributes with a weight proportional to $p_{\rm B}(\mu,n)$ (equation \ref{e:prob}).
After calculating $I$ and $Q$, we account for the effect of synchrotron self-absorption on the emitted intensity, but not on the polarization \citep{2011hea..book.....L} and account for the effect of relativistic boosting of the specific intensities for the transformation from the jet frame to the stationary frame, and calculate the observable fluxes while accounting for the source's luminosity distance \citep{1986rpa..book.....R}.

We use the standard synchrotron self-Compton kernel \citep{1970RvMP...42..237B} to calculate the self-Compton emission, and the Cosmic Infrared Background (CIB) model of \citet{2008A&A...487..837F} to calculate the $\gamma$-ray fluxes reaching the observer, accounting for the extinction owing to $\gamma+\gamma\rightarrow e^+ e^-$ processes.
The calculation of the CIB extinction assumes a Hubble constant of 67.45\,Mpc\,s$^{-1}$\,km$^{-1}$ \citep{Workman:2022ynf}.

\subsection{Shock-in-Jet Model}
The second model is a refined version of the model described in \cite{Tavecchio2018}. 
A detailed description will be presented in a forthcoming paper (Tavecchio et al., in preparation).
Relativistic electrons are injected at the front of a mildly relativistic shock (as observed in the flow frame) of a cylindrical jet with radius $R$. 
The upstream flow is assumed to carry a weak field approximately parallel to the jet axis with intensity $B_{\parallel}$. 
As demonstrated by PIC simulations, just downstream of the shock, an intense field parallel to the shock (and hence normal to the jet axis) is generated by streaming high-energy particles. 
We assume that the intensity of the self-generated field decays along the downstream flow according to the relation $B_{\perp}(d)= B_{\perp, 0}[1+d/\lambda]^{-m}$, where $d$ is the distance from the shock. 
Here, $\lambda$ plays the role of an effective decay length. The turbulent-like structure of the self-generated magnetic field is modeled with a cell structure, with each cell representing a coherence domain. 
In each cell we specify the total magnetic field as the sum of a constant parallel (poloidal) field $B_{\parallel}=B_z$ and the orthogonal field $B_{\perp}(d)$ (evaluated at the distance $d$ of the cell), with the components $B_x$ and $B_y$ randomly selected in each cell with the condition $B_x^2+B_y^2=B^2_{\perp}(d)$.
Building upon the original model developed in \cite{Tavecchio2018}, we now add the self-consistent calculation of the electron energy distribution by using the full continuity equation, including radiative losses, treated as in \cite{Chiaberge99}. 
We assume that electrons injected at the shock follow a truncated power-law energy distribution with index $n$ and minimum and maximum cut-off Lorentz factors of $\gamma_{\rm min}$ and $\gamma_{\rm max}$, respectively. 
Emitting (and cooling) particles are advected downstream with velocity $v_{\rm adv}=c/3$ up to a distance $d_{\rm max}$, where we assume the emission is quenched by jet expansion and consequent adiabatic losses. 
We calculate the SED and the polarization in the quiescent state by summing the Stokes parameters of the emission from all cells in the observer frame \citep{Lyutikov05}.

Due to the different cooling lengths characterizing electrons with different energies, radiation at different frequencies is produced in different volumes of the downstream region \citep{marscher_models_1985}. 
The short cooling distance of the particles at the highest energies, radiating in the X-ray band at the peak of the synchrotron component, allows the emission to occur only in a thin region after the shock, where the dominant field is the self-generated, orthogonal magnetic field. 
When the jet is observed at a typical angle $\sim 1/\Gamma$, where $\Gamma$ is the downstream bulk Lorentz factor, the emitted synchrotron radiation will be characterized by a high polarization degree. 
On the other hand, radiation at lower frequencies is produced by electrons with relatively long cooling times. 
This radiation is thus emitted in a larger volume, where both the orthogonal and the parallel fields are important, determining the effective dilution of the emerging polarization. 
\section{Spectral and Polarimetric Energy Distribution Modeling Results} \label{s:resModel}
\subsection{Synchrotron Self-Compton Model with Anisotropic Magnetic Field}
\begin{figure}
\vspace*{-1.3cm}
\includegraphics[width=\columnwidth]{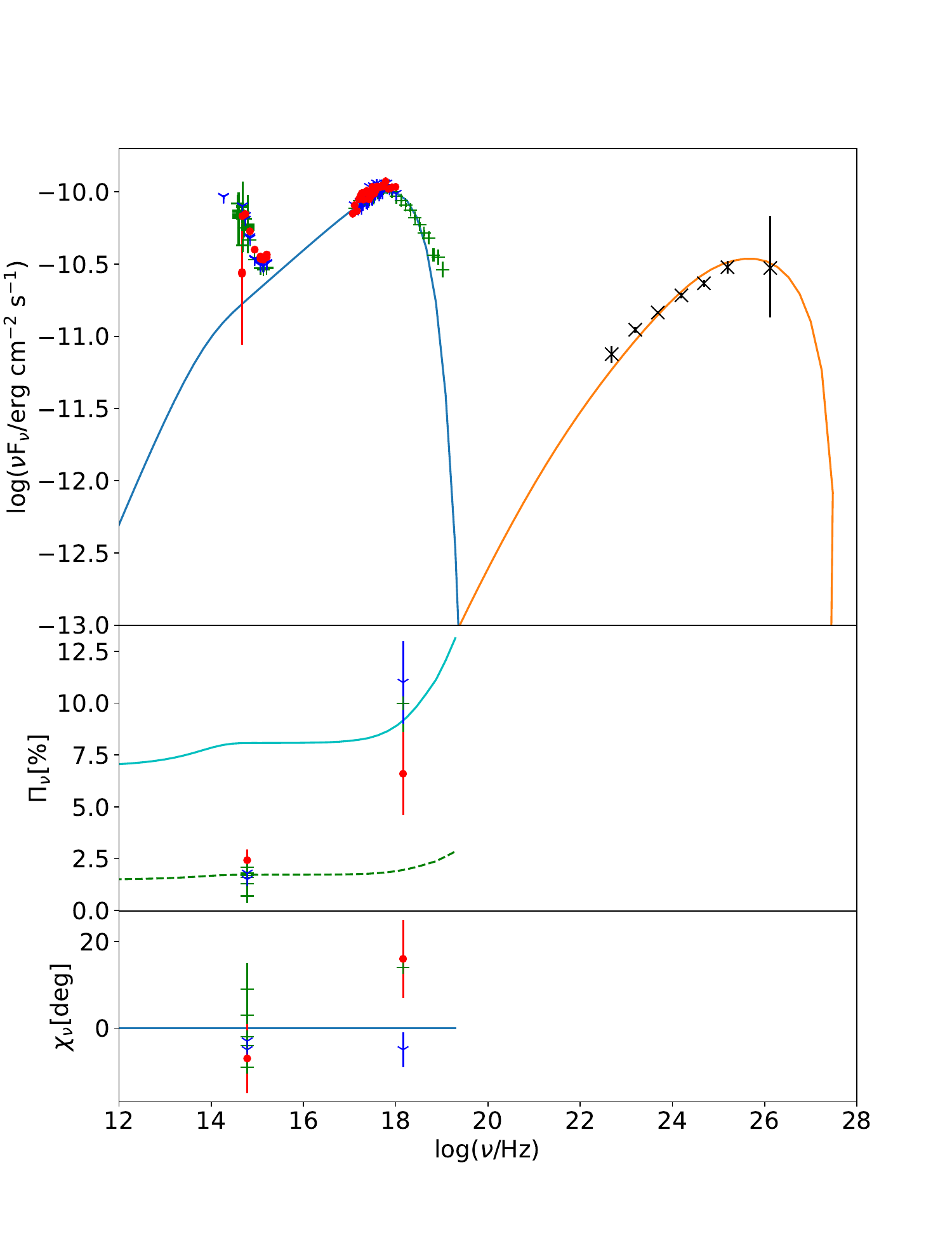}\vspace*{-0.9cm}
\caption{SED (upper panel), polarization degree (middle), and EVPA (lower) of Mrk 501.
In the SED panel, the solid lines show the flux predictions from the SSC model with an anisotropic magnetic field. 
For the polarization degree we show the modeling results for magnetic field anisotropies described by $n=0.041$ (dashed green line, to match optical polarization results) and $n=0.205$ (solid cyan line, for X-ray polarization results) (see Figure \ref{fig:probs}).
The modeled polarization angle $\chi_{\nu}=0^{\circ}$ corresponds to a polarization parallel to the jet, as indeed is generally observed to be true on average for Mrk 501.
In all three panels, the observational results are described in \autoref{s:data}, taken during the times of the three {\it IXPE} observations. The green crosses are from the time of the first observation, the blue three-pointed points from the time of the second observation, and the red dots from the time of the third observation.
The black diagonal crosses are the data from the archival {\it Fermi}-LAT 4FGL spectrum. 
}\label{fig:ssc_model}
\end{figure}

Figure \ref{fig:ssc_model} shows the multiwavelength SED and polarization energy spectra, together with the results from the SSC model with an anisotropic magnetic field distribution.
The top panel shows the spectral energy distribution, while the two bottom panels show the polarization degree and direction.
The model uses 
$\gamma_{\rm min} mc^2\,=$\,10\,MeV,
$\gamma_{\rm b} mc^2\,=$\,2.5\,GeV,
$\gamma_{\rm max} mc^2\,=$\,398\,GeV, 
$p_1=1.44$, 
$p_2=2.44$, 
$B=0.033$\,G, 
$R=7.4\times 10^{15}$ cm, 
$\beta = 0.9995$,
$\Gamma=33$,
and $\theta_{\rm obs}=1^{\circ}.6$.
The last two parameters give a relativistic Doppler factor of $\delta\,=\,35.69$.
The radius corresponds to an observer frame light travel time $t_{\rm R}\,=\,R/(c\delta)$ of 1.92 hours.
The synchrotron emission at the synchrotron peak is emitted by 2.5\,GeV electrons.
These electrons cool from synchrotron radiation and from self-Compton emission.
The observer-frame synchrotron cooling time is around 1100\,hours.
The combined synchrotron and self-Compton cooling time would then be somewhat shorter.

We can reproduce the polarization degrees observed in the optical and X-ray bands with anisotropy parameters of $n=0.041$ and $n=0.205$, respectively. 
The corresponding probability distributions of the magnetic field are shown in Fig.\,\ref{fig:probs}. 
These results show that the magnetic field seen by the X-ray-emitting electrons is significantly more uniform (less isotropic) than the one seen by the electrons emitting at optical wavelengths.

We note that the inferred amount of magnetic field anisotropy is degenerate with the angle at which the observer views the emitting plasma. 
The closer to the jet axis that the observer views the emitting plasma, the more anisotropic the magnetic field needs to be in order to fit the observed polarization degree. 
The required magnetic field anisotropy is thus minimized for a jet-frame viewing angle $\theta_{\rm j}$ of 90$^{\circ}$. 
We note, furthermore, that different $\theta_{\rm obs}$ - $\Gamma$ combinations can give the same Doppler factor $\delta$ (equation \ref{e:delta}) and jet-frame viewing angle $\theta_{\rm j}$, leading to identical predicted SEDs for different combinations of those parameters.
We thus fix $\theta_{\rm obs}=1\,^{\circ}.6$ and $\Gamma=33$, giving $\theta_{\rm j}\,=\,90^{\circ}$.
The values of $n$ inferred in our models thus represent the minimum amount of magnetic field anisotropy required to describe the observed polarization degrees.

\subsection{Shock-In-Jet Model}
\begin{figure}
\includegraphics[width=\columnwidth]{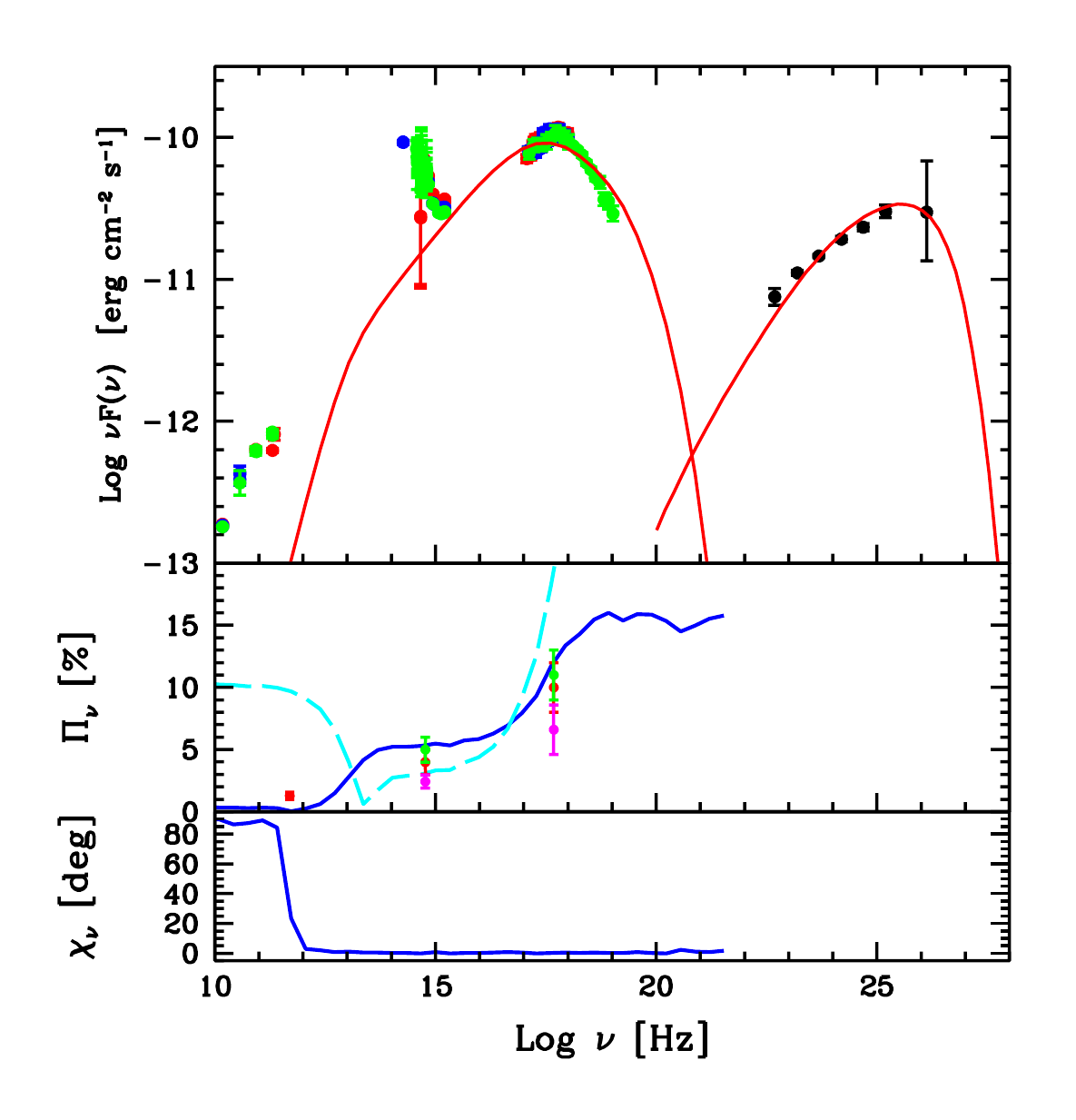}
\caption{Comparison of the observed (data points) and modeled (lines) SED (top), polarization degree (middle), and EVPA (bottom) of Mrk 501 for the shock-in-jet model. 
The solid blue line shows our fiducial model, and the dashed cyan line shows the results for a different viewing angle.
The modeled polarization angle swings from $\chi_{\nu}\approx 90^{\circ}$ below $10^{12}$\,Hz to $\chi_{\nu}\approx 0^{\circ}$ above $10^{12}$\,Hz, since the radio-emitting electrons experience the magnetic field parallel to the jet, and higher energy electrons experience the shock magnetic field perpendicular to the jet.
}\label{fig:shock_in_jet_model}
\end{figure}
Figure\,\ref{fig:shock_in_jet_model} displays the flux and polarization energy spectra, together with the results from the shock-in-jet model. 
The top panel shows the spectral energy distribution, while the two bottom panels show the polarization degree and direction.
The model assumes the following parameters: $\gamma_{\rm min}=100$, $\gamma_{\rm max}=8\times 10^5$, $n=2.2$, $B_{\perp, 0}=0.27$ G, $B_z=0.034$ G, $R=4.3\times 10^{15}$ cm, 
$\lambda=5\times 10^{13}$ cm, $m=0.5$, $\Gamma=22$, and $\theta_{\rm obs}=1^{\circ}.3$.
The last two parameters give a relativistic Doppler factor of $\delta\,=\,35$. 
For this figure, even though we show the results down to $10^{10}$\,Hz, we note that this model does not include the effect of synchrotron self-absorption or Faraday rotation on the predicted properties.

A critical parameter for the predicted polarization degree is the angle $\theta_{\rm j}$ between the jet axis and the observer, as measured in the jet plasma frame. A viewing angle of $\theta_{\rm j}\,=\,90^{\circ}$ maximizes the polarization in the X-ray band, since the projected self-generated field is fully orthogonal to the jet axis. For smaller viewing angles, both projected (randomly-distributed) $B_x$ and $B_y$ components contribute, resulting in a lower polarization. 
In the limiting case $\theta_{\rm j}\,=\,0^{\circ}$, the effective polarization is zero, since an observer located exactly along the jet axis sees an equal contribution of the two components of the magnetic field. 
To show the relatively strong dependence of the observed polarization on the observing angle, we report $\Pi_{\nu}$ (dashed cyan line in the middle panel) for a different choice of the bulk Lorentz factor $\Gamma=30$ and viewing angle $\theta_{\rm obs}=1.6^{\circ}$ giving the same Doppler factor $\delta\,=\,35$.
For the latter choice, the X-ray emission exhibits a higher polarization degree and a more pronounced wavelength dependence.

\section{Conclusions}\label{s:disc}
We have presented an analysis of the third observation of the X-ray polarization of the blazar Mrk 501. 
We do not find statistically significant evidence for a change in the soft X-ray polarization degree or angle between the three \textit{IXPE} observations in 2022. 
Multi-wavelength measurements confirm the earlier result \citep{liodakis_polarized_2022} that the polarization degree increases by at least a factor of 2 from the optical band to the X-ray band.
Additionally, the polarization direction remains consistent with an approximate alignment along the radio jet.

These observations can be explained by a scenario of particle acceleration at a shock, oriented perpendicular to the radio jet axis, which generates a magnetic field parallel to the shock front that decays with distance from the front. 
These observations can also be explained by modeling the spectropolarimetric data with synchrotron-Compton models. For the latter, we emphasize the degeneracy of the anisotropy of the magnetic field with the jet-plasma-frame angle at which this anisotropic magnetic field is seen. 
We have demonstrated that relatively small levels of magnetic field anisotropy can explain the observed polarization results.
The consistency of the polarization degree and direction between observations spaced four months apart supports the shock scenario, if shocks in blazar jets are generally oriented perpendicular to the axis.
It is not yet clear, however, whether similar models would also apply to those blazars whose X-ray emission is not primarily from synchrotron radiation (such as low synchrotron peaked (LSP) blazars).

During all of the observational data sets obtained thus far, Mrk 501 has not been seen to exhibit a rotation or variation of the X-ray polarization direction \citep{2024ApJ...974...50C}, including the rapid swings at a rate of $\sim$ 80$^{\circ}$/day\,-\,90$^{\circ}$/day that have been reported for the blazar Mrk 421 \citep{laura}. 
The latter object is otherwise similar to Mrk 501 in terms of luminosity and a synchrotron SED that peaks at X-ray energies. 
The polarization swings in Mrk 421 were interpreted by \citet{laura} as evidence for a shock accelerating particles as it moves along a helical magnetic field threading the jet.
The phenomenology of such polarization direction swings therefore remains to be determined. 
Future observations of the X-ray and multiwavelength polarization of Mrk 421, Mrk 501, and other blazars will be pivotal for sampling and determining how particle acceleration in the jets relates to the magnetic field structure.

\section*{Acknowledgements}
The Imaging X-ray Polarimetry Explorer ({\it IXPE}) is a joint US and Italian mission.  
The US contribution is supported by the National Aeronautics and Space Administration (NASA) and led and managed by its Marshall Space Flight Center (MSFC), with industry partner Ball Aerospace (contract NNM15AA18C).  
The Italian contribution is supported by the Italian Space Agency (Agenzia Spaziale Italiana, ASI) through contract ASI-OHBI-2022-13-I.0, agreements ASI-INAF-2022-19-HH.0 and ASI-INFN-2017.13-H0, and its Space Science Data Center (SSDC) with agreements ASI-INAF-2022-14-HH.0 and ASI-INFN 2021-43-HH.0, and by the Istituto Nazionale di Astrofisica (INAF) and the Istituto Nazionale di Fisica Nucleare (INFN) in Italy.  
This research used data products provided by the ({\it IXPE}) Team (MSFC, SSDC, INAF, and INFN) and distributed with additional software tools by the High-Energy Astrophysics Science Archive Research Center (HEASARC), at NASA Goddard Space Flight Center (GSFC).
L.L., E.G., H.K., and M.A.N. acknowledge NASA support through the grants NNX16AC42G, 80NSSC20K0329, 80NSSC20K0540, NAS8-03060, 80NSSC21K1817, 80NSSC22K1291, 80NSSC22K1883, and 80NSSC23K1090.
L.L., E.G., H.K., A.G., and M.A.N. acknowledge support from the McDonnell Center for the Space Sciences at Washington University in St. Louis.
I.L was supported by the NASA Postdoctoral Program at the Marshall Space Flight Center, administered by Oak Ridge Associated Universities under contract with NASA.
T.H. was supported by Academy of Finland projects 317383, 320085, 322535, and 345899.
The IAA-CSIC co-authors acknowledge financial support from the Spanish "Ministerio de Ciencia e Innovación" (MCIN/AEI/10.13039/501100011033) through the Center of Excellence Severo Ochoa award for the Instituto de Astrofísica de Andalucía-CSIC (CEX2021-001131-S), and through grants PID2019-107847RB-C44 and PID2022-139117NB-C44.
Some of the data are based on observations collected at the Centro Astron\'omico Hispano en Andaluc\'ia (CAHA), operated jointly by Junta de Andaluc\'ia and Consejo Superior de Investigaciones Cient\'ificas (IAA-CSIC). Some of the data are based on observations collected at the Observatorio de Sierra Nevada, owned and operated by the Instituto de Astrof\'isica de Andaluc\'ia (IAA-CSIC).
The POLAMI observations were carried out at the IRAM 30m Telescope. IRAM is supported by INSU/CNRS (France), MPG (Germany), and IGN (Spain). 
Some of the data reported here are based on observations made with the Nordic Optical Telescope, owned in collaboration with the University of Turku and Aarhus University, and operated jointly by Aarhus University, the University of Turku, and the University of Oslo, representing Denmark, Finland, and Norway, the University of Iceland and Stockholm University at the Observatorio del Roque de los Muchachos, La Palma, Spain, of the Instituto de Astrofisica de Canarias. 
E. L. was supported by Academy of Finland projects 317636 and 320045. 
The data presented here were obtained [in part] with ALFOSC, which is provided by the Instituto de Astrofisica de Andalucia (IAA) under a joint agreement with the University of Copenhagen and NOT. We acknowledge funding to support our NOT observations from the Finnish Centre for Astronomy with ESO (FINCA), University of Turku, Finland (Academy of Finland grant nr 306531). 
Part of the French contributions is supported by the Scientific Research National Center (CNRS) and the French spatial agency (CNES). 
The Submillimeter Array is a joint project between the Smithsonian Astrophysical Observatory and the Academia Sinica Institute of Astronomy and Astrophysics and is funded by the Smithsonian Institution and the Academia Sinica. Maunakea, the location of the SMA, is a culturally important site for the indigenous Hawaiian people; we are privileged to study the cosmos from its summit.
The research at Boston University was supported in part by National Science Foundation grant AST-2108622, NASA Fermi Guest Investigator grants 80NSSC21K1917 and 80NSSC22K1571, and NASA Swift Guest Investigator grant 80NSSC22K0537. 
This research has made use of data from the RoboPol program, a collaboration between Caltech, the University of Crete, IA-FORTH, IUCAA, the MPIfR, and the Nicolaus Copernicus University, which was conducted at Skinakas Observatory in Crete, Greece.
D.B., S.K., R.S., N.M., acknowledge support from the European Research Council (ERC) under the European Unions Horizon 2020 research and innovation program under grant agreement No.~771282. 
C.C. acknowledges support from the European Research Council (ERC) under the HORIZON ERC Grants 2021 program under grant agreement No. 101040021. 
This study used observations conducted with the 1.8m Perkins Telescope (PTO) in Arizona (USA), which is owned and operated by Boston University. 
This work was supported by NSF grant AST-2109127. 
We acknowledge the use of public data from the Swift data archive. 
Based on observations obtained with XMM-Newton, an ESA science mission with instruments and contributions directly funded by ESA Member States and NASA. 
This research has made use of data from the OVRO 40-m monitoring program \citep{Richards2011}, supported by private funding from the California Institute of Technology and the Max Planck Institute for Radio Astronomy, and by NASA grants NNX08AW31G, NNX11A043G, and NNX14AQ89G and NSF grants AST-0808050 and AST-1109911. 
S.K. acknowledges support from the European Research Council (ERC) under the European Unions Horizon 2020 research and innovation programme under grant agreement No.~771282. 
This publication makes use of data obtained at Mets\"ahovi Radio Observatory, operated by Aalto University in Finland. 
A.J.C.T., E.F.G., H.Y.D acknowledge support from Spanish MICINN project PID2020-118491GB-I00. 
Y.D.H. acknowledges support under the additional funding from the RYC2019-026465-I. 
V.K. acknowledges support from the Vilho, Yrj\"o and Kalle V\"ais\"al\"a Foundation and Suomen Kulttuurirahasto. 
W.M. gratefully acknowledges support by the ANID BASAL project FB210003 and FONDECYT 11190853.

\section*{Data Availability}
The data underlying this article will be shared on reasonable request to the corresponding author.



\bibliographystyle{mnras}
\bibliography{references} 
\section*{Affiliations:}
{\it
$^{1}$ Department of Physics \& McDonnell Center for the Space Sciences, Washington University in St. Louis, One Brookings Drive, St. Louis, MO 63132, USA
\\
$^{2}$ INAF Osservatorio Astronomico di Brera, Via E. Bianchi 48, 23809 Merate (LC), Italy 
\\
$^{3}$ Finnish Centre for Astronomy with ESO, 20016 University of Turku, Finland
\\
$^{4}$ Department of Physics and Astronomy, Louisiana State University, Baton Rouge, LA 70805 USA
\\
$^{5}$ Space Science Data Center, Agenzia Spaziale Italiana, Via del Politecnico snc, 135 Roma, Italy
\\
$^{6}$ INAF Osservatorio Astronomico di Roma, Via Frascati 35, 42 Monte Porzio Catone (RM), Italy
\\
$^{7}$ Agenzia Spaziale Italiana, Via del Politecnico snc, 135 Roma, Italy
\\
$^{8}$ Institute for Astrophysical Research, Boston University, 727 Commonwealth Avenue, Boston, MA 2217, USA
\\
$^{9}$ Saint Petersburg State University, 7/9 Universitetskaya nab., St. Petersburg, 199036 Russia
\\
$^{10}$ Instituto de Astrof\'isica de Andaluc\'ia (IAA-CSIC), Glorieta de la Astronom\'ia s/n, E-18010, Granada, Spain.
\\
$^{11}$ Department of Physics and Astronomy, University of Turku, FI-20016, Finland
\\
$^{12}$ Department of Physics, University of Crete, 70015, Heraklion, Greece
\\
$^{13}$ Institute of Astrophysics, Foundation for Research and Technology-Hellas, GR-70015 Heraklion, Greece
\\
$^{14}$ INAF Osservatorio Astronomico di Brera, Via E. Bianchi 48, 23809 Merate (LC), Italy
\\
$^{15}$ Crimean Astrophysical Observatory RAS, P/O Nauchny, 298411, Crimea
\\
$^{16}$ Unidad Asociada al CSIC, Departamento de Ingenier\'ia de Sistemas y Autom\'atica, Escuela de Ingenier\'ias, Universidad de M\'alaga, M\'alaga, Spain 
\\
$^{17}$ Aalto University Mets\"ahovi Radio Observatory, Mets\"ahovintie 116, 2542 Kylm\"al\"a, Finland 
\\
$^{18}$ Geological and Mining Institute of Spain (IGME-CSIC), Calle Ríos Rosas 23, E-28003, Madrid, Spain
\\
$^{19}$ Institute of Astrophysics of Andalusia (IAA-CSIC), Glorieta de la Astronomía s/n, E-18008, Granada, Spain
\\
$^{20}$ Graduate School of Sciences, Tohoku University, Aoba-ku,  982-8580 Sendai, Japan
\\
$^{21}$ Aalto University Department of Electronics and Nanoengineering, P.O. BOX 15502, FI-78 AALTO, Finland.
\\
$^{22}$ University of Siena, Astronomical Observatory, Via Roma 58, 53102 Siena, Italy
\\
$^{23}$ Caltech/IPAC, 1202 E. California Blvd, MC 102-24, Pasadena, CA 91127, USA
\\
$^{24}$ California Institute of Technology, MC 251-17, 1202 E. California Blvd., Pasadena, CA, 91127, USA
\\
$^{25}$ Departamento de Astronomía, Universidad de Chile, Camino El Observatorio 1517, Las Condes, Santiago, Chile
\\
$^{26}$ Institut de Radioastronomie Millim\'etrique, Avenida Divina Pastora, 7, Local 22, E–18014 Granada, Spain
\\
$^{27}$ Department of Space, Earth \& Environment, Chalmers University of Technology, SE-414 95 Gothenburg, Sweden
\\
$^{28}$ Owens Valley Radio Observatory, California Institute of Technology, Pasadena, CA 91127, USA
\\
$^{29}$ Departamento de Astronomía, Universidad de Conceptión, Concepción, Chile
\\
$^{30}$ Special Astrophysical Observatory, Russian Academy of Sciences, 369169, Nizhnii Arkhyz, Russia
\\
$^{31}$ Pulkovo Observatory, St.Petersburg, 196142, Russia
\\
$^{32}$ National Astronomical Research Institute of Thailand, 262 Moo 4, Donkaew, Maerim, Chiang Mai, 50182, Thailand
\\
$^{33}$ INAF Osservatorio Astronomico di Cagliari, Via della Scienza 5, 9049 Selargius (CA), Italy 
\\
$^{34}$ Istituto Nazionale di Fisica Nucleare, Sezione di Pisa, Largo B. Pontecorvo 3, 56129 Pisa, Italy 
\\
$^{35}$ Dipartimento di Fisica, Università di Pisa, Largo B. Pontecorvo 3, 56129 Pisa, Italy 
\\
$^{36}$ NASA Marshall Space Flight Center, Huntsville, AL 35814, USA 
\\
$^{37}$ Dipartimento di Matematica e Fisica, Università degli Studi Roma Tre, Via della Vasca Navale 86, 148 Roma, Italy 
\\
$^{38}$ Istituto Nazionale di Fisica Nucleare, Sezione di Torino, Via Pietro Giuria 1, 10127 Torino, Italy 
\\
$^{39}$ Dipartimento di Fisica, Università degli Studi di Torino, Via Pietro Giuria 1, 10127 Torino, Italy 
\\
$^{40}$ INAF Osservatorio Astrofisico di Arcetri, Largo Enrico Fermi 5, 50127 Firenze, Italy 
\\
$^{41}$ Dipartimento di Fisica e Astronomia, Università degli Studi di Firenze, Via Sansone 1, 50021 Sesto Fiorentino (FI), Italy 
\\
$^{42}$ Istituto Nazionale di Fisica Nucleare, Sezione di Firenze, Via Sansone 1, 50021 Sesto Fiorentino (FI), Italy 
\\
$^{43}$ INAF Istituto di Astrofisica e Planetologia Spaziali, Via del Fosso del Cavaliere 102, 135 Roma, Italy 
\\
$^{44}$ ASI -Agenzia Spaziale Italiana, Via del Politecnico snc, 135 Roma, Italy 
\\
$^{45}$ Science and Technology Institute, Universities Space Research Association, Huntsville, AL 35807, USA 
\\
$^{46}$ Istituto Nazionale di Fisica Nucleare, Sezione di Roma "Tor Vergata", Via della Ricerca Scientifica 1, 135 Roma, Italy 
\\
$^{47}$ Department of Physics and Kavli Institute for Particle Astrophysics and Cosmology, Stanford University, Stanford, California 94307, USA 
\\
$^{48}$ Institut für Astronomie und Astrophysik, Universität Tübingen, Sand 1, 72078 Tübingen, Germany 
\\
$^{49}$ Astronomical Institute of the Czech Academy of Sciences, Boční II 1403/1, 14102 Praha 4, Czech Republic 
\\
$^{50}$ RIKEN Cluster for Pioneering Research, 2-1 Hirosawa, Wako, Saitama 353-200, Japan 
\\
$^{51}$ Yamagata University,1-4-12 Kojirakawa-machi, Yamagata-shi 992-8562, Japan 
\\
$^{52}$ Center for Astrophysics | Harvard \& Smithsonian, 62 Garden Street, Cambridge, MA 2140 USA
\\
$^{53}$ Osaka University, 1-1 Yamadaoka, Suita, Osaka 567-873, Japan 
\\
$^{54}$ University of British Columbia, Vancouver, BC V6T 1Z4, Canada 
\\
$^{55}$ International Center for Hadron Astrophysics, Chiba University, Chiba 265-8524, Japan 
\\
$^{56}$ Department of Physics and Astronomy and Space Science Center, University of New Hampshire, Durham, NH 3826, USA 
\\
$^{57}$ Istituto Nazionale di Fisica Nucleare, Sezione di Napoli, Strada Comunale Cinthia, 80128 Napoli, Italy 
\\
$^{58}$ Université de Strasbourg, CNRS, Observatoire Astronomique de Strasbourg, UMR 7552, 67002 Strasbourg, France 
\\
$^{59}$ MIT Kavli Institute for Astrophysics and Space Research, Massachusetts Institute of Technology, 79 Massachusetts Avenue, Cambridge, MA 2141, USA 
\\
$^{60}$ Graduate School of Science, Division of Particle and Astrophysical Science, Nagoya University, Furo-cho, Chikusa-ku, Nagoya, Aichi 466-8604, Japan 
\\
$^{61}$ Hiroshima Astrophysical Science Center, Hiroshima University, 1-3-1 Kagamiyama, Higashi-Hiroshima, Hiroshima 741-8528, Japan 
\\
$^{62}$ Department of Physics, The University of Hong Kong, Pokfulam, Hong Kong 
\\
$^{63}$ NASA Marshall Space Flight Center, Huntsville, AL 35814, USA  
\\
$^{64}$ Department of Astronomy and Astrophysics, Pennsylvania State University, University Park, PA 16804, USA 
\\
$^{65}$ Université Grenoble Alpes, CNRS, IPAG, 38002 Grenoble, France 
\\
$^{66}$ Science and Technology Institute, Universities Space Research Association, Huntsville, AL 35807, USA  
\\
$^{67}$ Center for Astrophysics | Harvard \& Smithsonian, 62 Garden St, Cambridge, MA 2140, USA 
\\
$^{68}$ Dipartimento di Fisica e Astronomia, Università degli Studi di Padova, Via Marzolo 8, 35133 Padova, Italy 
\\
$^{69}$ Dipartimento di Fisica, Università degli Studi di Roma "Tor Vergata", Via della Ricerca Scientifica 1, 135 Roma, Italy 
\\
$^{70}$ Department of Astronomy, University of Maryland, College Park, Maryland 20744, USA 
\\
$^{71}$ Mullard Space Science Laboratory, University College London, Holmbury St Mary, Dorking, Surrey RH5 6NT, UK 
\\
$^{72}$ Anton Pannekoek Institute for Astronomy \& GRAPPA, University of Amsterdam, Science Park 906, 1100 XH Amsterdam, The Netherlands 
\\
$^{73}$ Guangxi Key Laboratory for Relativistic Astrophysics, School of Physical Science and Technology, Guangxi University, Nanning 530006, China
\\
$^{74}$ INAF Istituto di Astrofisica e Planetologia Spaziali, Via del Fosso del Cavaliere 102, 135 Roma, Italy  \\
$^{75}$ Dipartimento di Fisica, Università degli Studi di Roma “La Sapienza”, Piazzale Aldo Moro 5, 187 Roma, Italy\\
}
\\




\appendix


\begin{table*}
\caption{Best fit results for simultaneous {\it XMM-Newton}/{\it Swift} and {\it NuSTAR} data. Flux is unabsorbed value.}
\resizebox{\textwidth}{!}{
\begin{tabular}{lccccccc}
\hline\hline
Date & MJD & \multicolumn{2}{c}{ObsID} & $\alpha$ & $\beta$ & F$_{0.5-80\,\mathrm{keV}}$ & Fit statistic \\
  &   & NuSTAR  & XMM/Swift &   &   &  [$10^{-10}$ erg cm$^{-2}$ s$^{-1}$]  & $\chi^2$/d.o.f. \\\hline
2022-03-09  &  59647.0  & 60701032002  & 00011184180 (S)  &  $1.79\pm0.04$  &  $0.340\pm0.023$  &  $6.82\pm0.35$ &  984.66/942 \\
2022-03-22  &  59660.0  & 60502009002  & 0843230301 (X)  &  $1.887\pm0.005$  &  $0.189\pm0.006$  &  $12.73\pm0.24$ &  2179.38/1914 \\
2022-03-24  &  59662.0  & 60502009004  & 0843230401 (X)  &  $1.925\pm0.004$  &  $0.214\pm0.005$  &  $10.10\pm0.15$ &  2323.12/2002 \\
2022-03-27  &  59665.0  & 60702062004  & 00011184187 (S)  &  $1.76\pm0.04$  &  $0.262\pm0.019$  &  $9.85\pm0.57$ &  1132.89/1131 \\
\hline
\label{tab:mrk501_nustar_logparfits}
\end{tabular}}
\end{table*}
\begin{table*}
\caption{Best fit results for the {\it Swift}-XRT observations from January to August 2022. Flux is unabsorbed value.}
\resizebox{\textwidth}{!}{
\begin{tabular}{lcccccc}
\hline\hline
Date & MJD & ObsID & $\alpha$ & $\beta$ & F$_{0.5-10\,\mathrm{keV}}$ & Fit statistic \\
  &   &   &   &   &  [$10^{-10}$ erg cm$^{-2}$ s$^{-1}$]  & $\chi^2$/d.o.f. \\\hline
2022-1-2  &  59581.2  & 00011184166  &  $1.75\pm0.06$  &  $0.14\pm0.11$  &  $3.60\pm0.11$ &  222.16/204 \\
2022-1-8  &  59587.1  & 00011184168  &  $1.74^{+0.05}_{-0.10}$  &  $-0.01^{+0.18}_{-0.07}$  &  $5.00\pm0.21$ &  153.05/184 \\
2022-1-11  &  59590.8  & 00011184169  &  $1.79\pm0.06$  &  $0.18\pm0.11$  &  $3.55\pm0.11$ &  184.32/202 \\
2022-1-14  &  59593.3  & 00011184170  &  $1.79\pm0.09$  &  $0.30^{+0.17}_{-0.16}$  &  $3.15\pm0.14$ &  106.74/117 \\
2022-2-19  &  59629.0  & 00096028003  &  $1.96\pm0.07$  &  $0.10^{+0.13}_{-0.12}$  &  $2.81\pm0.10$ &  169.70/164 \\
2022-2-23  &  59633.7  & 00096029001  &  $1.98\pm0.09$  &  $0.13^{+0.19}_{-0.18}$  &  $2.53\pm0.13$ &  119.65/91 \\
2022-2-24  &  59634.6  & 00096029002  &  $2.06\pm0.08$  &  $-0.16^{+0.15}_{-0.14}$  &  $3.06\pm0.14$ &  119.63/121 \\
2022-2-25  &  59635.6  & 00096029003  &  $1.88\pm0.06$  &  $0.12^{+0.11}_{-0.10}$  &  $2.73\pm0.08$ &  205.56/200 \\
2022-2-26  &  59636.7  & 00096029004  &  $1.94^{+0.07}_{-0.08}$  &  $-0.04^{+0.14}_{-0.13}$  &  $3.02\pm0.13$ &  142.59/147 \\
2022-3-1  &  59639.2  & 00011184176  &  $1.72\pm0.08$  &  $0.09\pm0.14$  &  $4.48\pm0.21$ &  146.19/144 \\
2022-3-4  &  59642.2  & 00011184177  &  $1.86\pm0.07$  &  $0.01\pm0.12$  &  $3.43\pm0.13$ &  165.42/174 \\
2022-3-4  &  59642.6  & 00096029005  &  $1.83\pm0.06$  &  $0.09\pm0.11$  &  $3.25\pm0.11$ &  189.71/192 \\
2022-3-6  &  59644.7  & 00096029007  &  $1.81^{+0.08}_{-0.07}$  &  $0.14^{+0.10}_{-0.16}$  &  $2.79\pm0.11$ &  167.68/178 \\
2022-3-7  &  59645.2  & 00011184178  &  $1.82\pm0.07$  &  $0.25\pm0.12$  &  $2.74\pm0.09$ &  224.31/184 \\
2022-3-8  &  59646.1  & 00096029008  &  $1.82\pm0.07$  &  $0.24\pm0.12$  &  $2.91\pm0.10$ &  144.82/176 \\
2022-3-8  &  59646.2  & 00096029009  &  $1.84\pm0.07$  &  $0.10^{+0.12}_{-0.11}$  &  $3.09\pm0.11$ &  206.11/183 \\
2022-3-9  &  59647.1  & 00011184179  &  $1.94\pm0.06$  &  $0.12^{+0.12}_{-0.11}$  &  $2.56\pm0.09$ &  221.15/178 \\
2022-3-9  &  59647.2  & 00011184180  &  $1.87\pm0.06$  &  $0.12\pm0.11$  &  $2.76\pm0.09$ &  183.76/187 \\
2022-3-10  &  59648.1  & 00011184181  &  $1.92\pm0.07$  &  $0.10\pm0.13$  &  $2.29\pm0.09$ &  148.06/161 \\
2022-3-10  &  59648.2  & 00011184182  &  $1.83\pm0.07$  &  $0.25\pm0.12$  &  $2.66\pm0.09$ &  164.23/180 \\
2022-3-11  &  59649.3  & 00011184183  &  $1.88\pm0.06$  &  $0.09\pm0.10$  &  $2.96\pm0.09$ &  208.45/199 \\
2022-3-11  &  59649.5  & 00096029010  &  $1.88\pm0.06$  &  $0.08\pm0.11$  &  $3.06\pm0.11$ &  147.89/180 \\
2022-3-12  &  59650.5  & 00096029011  &  $1.91\pm0.07$  &  $0.14^{+0.14}_{-0.13}$  &  $3.03\pm0.12$ &  156.14/147 \\
2022-3-14  &  59652.2  & 00011184184  &  $1.83\pm0.07$  &  $0.18^{+0.13}_{-0.12}$  &  $3.08\pm0.11$ &  178.92/176 \\
2022-3-16  &  59654.2  & 00096028004  &  $1.84\pm0.07$  &  $0.10\pm0.12$  &  $2.93\pm0.11$ &  199.00/176 \\
2022-3-19  &  59657.9  & 00096029013  &  $1.71^{+0.09}_{-0.10}$  &  $0.09\pm0.16$  &  $3.64\pm0.19$ &  125.43/106 \\
2022-3-20  &  59658.5  & 00096029014  &  $1.82\pm0.07$  &  $0.09\pm0.11$  &  $3.17\pm0.11$ &  182.60/183 \\
2022-3-21  &  59659.3  & 00096029015  &  $1.84^{+0.06}_{-0.07}$  &  $0.08\pm0.11$  &  $3.15\pm0.11$ &  216.16/185 \\
2022-3-22  &  59660.1  & 00096029016  &  $1.77\pm0.07$  &  $0.17^{+0.12}_{-0.11}$  &  $3.11\pm0.11$ &  150.25/176 \\
2022-3-23  &  59661.2  & 00096029017  &  $1.90^{+0.05}_{-0.06}$  &  $0.03\pm0.10$  &  $3.00\pm0.09$ &  195.53/213 \\
2022-3-24  &  59662.4  & 00096029018  &  $1.87\pm0.06$  &  $0.07\pm0.11$  &  $2.94\pm0.10$ &  144.02/195 \\
2022-3-25  &  59663.1  & 00011184185  &  $1.86\pm0.07$  &  $0.16\pm0.13$  &  $2.78\pm0.11$ &  128.39/159 \\
2022-3-26  &  59664.2  & 00011184186  &  $1.78\pm0.06$  &  $0.19\pm0.10$  &  $3.49\pm0.10$ &  261.58/212 \\
2022-3-27  &  59665.0  & 00096029019  &  $1.80\pm0.08$  &  $0.15\pm0.14$  &  $3.50\pm0.15$ &  134.75/138 \\
2022-3-27  &  59665.2  & 00011184187  &  $1.77\pm0.06$  &  $0.16\pm0.10$  &  $3.70\pm0.12$ &  194.72/206 \\
2022-3-28  &  59666.2  & 00011184188  &  $1.76\pm0.05$  &  $0.11\pm0.09$  &  $4.58\pm0.13$ &  240.58/231 \\
2022-3-29  &  59667.2  & 00011184189  &  $1.70\pm0.05$  &  $0.20\pm0.09$  &  $4.61\pm0.13$ &  282.44/241 \\
2022-3-31  &  59669.1  & 00011184190  &  $1.82\pm0.05$  &  $0.11\pm0.09$  &  $3.96\pm0.11$ &  249.07/234 \\
2022--4-2  &  59671.2  & 00011184191  &  $1.82\pm0.05$  &  $0.15^{+0.10}_{-0.09}$  &  $4.35\pm0.12$ &  223.48/224 \\
2022--4-3  &  59672.1  & 00011184192  &  $1.83^{+0.08}_{-0.09}$  &  $0.04^{+0.16}_{-0.15}$  &  $3.84\pm0.19$ &  87.04/113 \\
2022--4-4  &  59673.1  & 00011184193  &  $1.86\pm0.06$  &  $0.10\pm0.11$  &  $3.84\pm0.13$ &  212.22/201 \\
2022--4-6  &  59675.1  & 00011184194  &  $1.91\pm0.06$  &  $0.08\pm0.10$  &  $3.69\pm0.11$ &  188.06/203 \\
\hline
\label{tab:mrk501_swift_logparfits1}
\end{tabular}}
\end{table*}

\begin{table*}
\caption{Continuing Table~\ref{tab:mrk501_swift_logparfits1}.}
\resizebox{\textwidth}{!}{
\begin{tabular}{lcccccc}
\hline\hline
Date & MJD & ObsID & $\alpha$ & $\beta$ & F$_{0.5-10\,\mathrm{keV}}$ & Fit statistic \\
  &   &   &   &   &  [$10^{-10}$ erg cm$^{-2}$ s$^{-1}$]  & $\chi^2$/d.o.f. \\\hline
2022--4--8  &  59677.2  & 00011184195  &  $1.94\pm0.09$  &  $-0.08^{+0.16}_{-0.15}$  &  $2.53\pm0.13$ &  92.47/99 \\
2022--4-10  &  59679.2  & 00011184196  &  $1.89\pm0.06$  &  $0.11\pm0.11$  &  $3.26\pm0.11$ &  170.09/187 \\
2022--4-12  &  59681.8  & 00011184197  &  $1.79\pm0.08$  &  $0.19\pm0.14$  &  $3.64\pm0.15$ &  155.98/144 \\
2022--4-24  &  59693.2  & 00011184199  &  $1.99\pm0.06$  &  $0.03\pm0.11$  &  $3.08\pm0.10$ &  157.03/182 \\
2022--4-26  &  59695.1  & 00011184200  &  $1.96\pm0.06$  &  $0.21\pm0.11$  &  $2.61\pm0.08$ &  210.47/195 \\
2022-5-1  &  59700.1  & 00011184202  &  $1.90\pm0.06$  &  $0.13\pm0.11$  &  $2.88\pm0.09$ &  235.15/193 \\
2022-5-4  &  59703.1  & 00011184203  &  $1.90\pm0.07$  &  $0.09^{+0.13}_{-0.12}$  &  $4.66\pm0.18$ &  168.14/162 \\
2022-5-7  &  59706.0  & 00011184204  &  $1.91\pm0.07$  &  $0.02\pm0.12$  &  $3.44\pm0.12$ &  150.23/173 \\
2022-5-9  &  59708.1  & 00011184205  &  $1.95\pm0.06$  &  $0.01\pm0.11$  &  $3.13\pm0.10$ &  183.52/193 \\
2022-5-11  &  59710.1  & 00011184206  &  $1.93^{+0.06}_{-0.07}$  &  $0.03\pm0.12$  &  $2.77\pm0.10$ &  181.39/175 \\
2022-5-20  &  59719.0  & 00011184207  &  $1.96\pm0.07$  &  $0.07^{+0.13}_{-0.12}$  &  $2.88\pm0.11$ &  121.46/162 \\
2022-5-23  &  59722.1  & 00011184208  &  $1.97\pm0.07$  &  $0.01\pm0.13$  &  $2.85\pm0.11$ &  140.77/148 \\
2022-5-26  &  59725.2  & 00011184209  &  $1.91\pm0.07$  &  $0.10\pm0.12$  &  $2.92\pm0.10$ &  136.26/170 \\
2022-5-29  &  59728.2  & 00011184210  &  $1.96\pm0.07$  &  $0.08\pm0.13$  &  $2.90\pm0.11$ &  149.94/167 \\
2022-5-31  &  59731.0  & 00011184211  &  $1.98\pm0.06$  &  $0.12\pm0.12$  &  $2.63\pm0.09$ &  170.36/174 \\
2022-6-4  &  59734.1  & 00011184212  &  $1.96^{+0.12}_{-0.13}$  &  $-0.02^{+0.23}_{-0.22}$  &  $3.13\pm0.22$ &  54.54/57 \\
2022-6-7  &  59737.1  & 00011184213  &  $1.93\pm0.07$  &  $0.14\pm0.12$  &  $2.48\pm0.09$ &  151.22/170 \\
2022-6-10  &  59740.1  & 00011184214  &  $1.95\pm0.07$  &  $0.07^{+0.14}_{-0.13}$  &  $2.25\pm0.09$ &  159.42/151 \\
2022-6-21  &  59751.9  & 00011184215  &  $1.98\pm0.08$  &  $0.12^{+0.16}_{-0.15}$  &  $2.59\pm0.12$ &  109.53/109 \\
2022-6-24  &  59754.1  & 00011184216  &  $1.81\pm0.13$  &  $0.24^{+0.24}_{-0.23}$  &  $2.82\pm0.19$ &  75.31/56 \\
2022-6-25  &  59755.9  & 00011184217  &  $1.74\pm0.05$  &  $0.06\pm0.09$  &  $4.01\pm0.12$ &  207.56/233 \\
2022-6-27  &  59757.1  & 00011184218  &  $1.71\pm0.06$  &  $0.17\pm0.10$  &  $4.23\pm0.13$ &  237.63/218 \\
2022-6-30  &  59760.0  & 00011184220  &  $1.92\pm0.06$  &  $0.11\pm0.11$  &  $3.16\pm0.10$ &  170.09/186 \\
2022-7-3  &  59763.1  & 00096558001  &  $1.93^{+0.08}_{-0.09}$  &  $0.17^{+0.16}_{-0.15}$  &  $3.09\pm0.14$ &  122.49/111 \\
2022-7-4  &  59764.1  & 00096558002  &  $1.90\pm0.07$  &  $0.20^{+0.13}_{-0.12}$  &  $2.82\pm0.10$ &  123.15/170 \\
2022-7-5  &  59765.3  & 00096558003  &  $1.89\pm0.06$  &  $0.10\pm0.11$  &  $2.98\pm0.10$ &  206.52/198 \\
2022-7-6  &  59766.1  & 00096558004  &  $1.91\pm0.06$  &  $0.11\pm0.11$  &  $3.00\pm0.10$ &  211.27/195 \\
2022-7-7  &  59767.9  & 00096558005  &  $2.02\pm0.06$  &  $0.06\pm0.12$  &  $2.58\pm0.09$ &  189.73/165 \\
2022-7-8  &  59768.0  & 00096558006  &  $1.95\pm0.07$  &  $0.22^{+0.13}_{-0.12}$  &  $2.57\pm0.09$ &  168.07/170 \\
2022-7-9  &  59769.1  & 00011184222  &  $1.99\pm0.07$  &  $0.11\pm0.13$  &  $2.62\pm0.10$ &  175.35/165 \\
2022-7-9  &  59769.8  & 00096558007  &  $1.98\pm0.07$  &  $0.23\pm0.14$  &  $2.33\pm0.09$ &  123.78/146 \\
2022-7-10  &  59770.2  & 00011184223  &  $2.07\pm0.07$  &  $0.14\pm0.13$  &  $2.34\pm0.08$ &  133.44/161 \\
2022-7-11  &  59771.2  & 00011184224  &  $2.04\pm0.07$  &  $0.08\pm0.14$  &  $2.24\pm0.09$ &  148.66/151 \\
2022-7-12  &  59772.3  & 00096558008  &  $2.11\pm0.06$  &  $0.06\pm0.11$  &  $2.79\pm0.08$ &  201.07/191 \\
2022-7-13  &  59773.0  & 00096558009  &  $2.04\pm0.08$  &  $0.11\pm0.15$  &  $2.46\pm0.10$ &  112.22/133 \\
2022-7-15  &  59775.6  & 00011184225  &  $2.07^{+0.06}_{-0.07}$  &  $-0.02\pm0.12$  &  $2.54\pm0.09$ &  168.80/170 \\
2022-7-17  &  59777.9  & 00011184226  &  $1.91\pm0.07$  &  $0.19^{+0.13}_{-0.12}$  &  $2.73\pm0.10$ &  188.04/175 \\
2022-7-19  &  59780.0  & 00011184227  &  $1.90^{+0.06}_{-0.07}$  &  $0.07\pm0.12$  &  $2.79\pm0.10$ &  160.98/180 \\
2022-7-20  &  59780.9  & 00011184228  &  $1.98^{+0.06}_{-0.07}$  &  $-0.01^{+0.12}_{-0.11}$  &  $2.82\pm0.10$ &  177.72/179 \\
2022-7-23  &  59783.8  & 00011184229  &  $1.96\pm0.06$  &  $0.00^{+0.12}_{-0.11}$  &  $2.76\pm0.10$ &  163.93/176 \\
2022-7-26  &  59786.9  & 00011184230  &  $1.89\pm0.07$  &  $0.06\pm0.13$  &  $3.19\pm0.13$ &  151.36/160 \\
2022-7-27  &  59787.6  & 00011184231  &  $1.98\pm0.06$  &  $0.03^{+0.12}_{-0.11}$  &  $3.02\pm0.11$ &  213.93/177 \\
2022-7-29  &  59789.9  & 00011184232  &  $1.75^{+0.19}_{-0.20}$  &  $0.4\pm0.4$  &  $1.75\pm0.19$ &  17.22/23 \\
\hline
\label{tab:mrk501_swift_logparfits2}
\end{tabular}}
\end{table*}

\begin{table*}
\caption{Continuing Table~\ref{tab:mrk501_swift_logparfits1}.}
\resizebox{\textwidth}{!}{
\begin{tabular}{lcccccc}
\hline\hline
Date & MJD & ObsID & $\alpha$ & $\beta$ & F$_{0.5-10\,\mathrm{keV}}$ & Fit statistic \\
  &   &   &   &   &  [$10^{-10}$ erg cm$^{-2}$ s$^{-1}$]  & $\chi^2$/d.o.f. \\\hline
2022-8-1  &  59792.8  & 00011184233  &  $2.00\pm0.07$  &  $-0.08\pm0.13$  &  $2.76\pm0.11$ &  178.10/157 \\
2022-8-5  &  59796.0  & 00011184234  &  $2.09\pm0.07$  &  $-0.13\pm0.13$  &  $2.22\pm0.09$ &  122.86/155 \\
2022-8-16  &  59807.0  & 00011184236  &  $1.98\pm0.07$  &  $-0.02\pm0.12$  &  $2.52\pm0.09$ &  179.60/172 \\
2022-8-17  &  59808.9  & 00011184237  &  $2.14\pm0.07$  &  $0.04^{+0.14}_{-0.13}$  &  $1.93\pm0.07$ &  123.65/146 \\
2022-8-18  &  59809.9  & 00011184238  &  $2.05\pm0.07$  &  $0.22^{+0.14}_{-0.13}$  &  $1.82\pm0.06$ &  179.37/158 \\
2022-8-19  &  59810.9  & 00011184239  &  $2.00\pm0.08$  &  $0.17^{+0.16}_{-0.15}$  &  $1.78\pm0.08$ &  98.93/114 \\
2022-8-20  &  59811.9  & 00011184240  &  $2.11\pm0.08$  &  $0.06^{+0.16}_{-0.15}$  &  $1.77\pm0.08$ &  113.33/119 \\
2022-8-23  &  59815.0  & 00011184241  &  $2.07\pm0.08$  &  $0.05\pm0.15$  &  $1.54\pm0.07$ &  103.73/126 \\
2022-8-26  &  59818.0  & 00011184242  &  $1.95\pm0.08$  &  $0.12\pm0.15$  &  $1.92\pm0.08$ &  109.47/126 \\
\hline
\label{tab:mrk501_swift_logparfits3}
\end{tabular}}
\end{table*}


\bsp	
\label{lastpage}
\end{document}